\documentclass[aip,graphicx]{revtex4-1}

\usepackage{amsmath}
\usepackage{amssymb}
\usepackage[latin9]{inputenc}
\usepackage{graphicx}
\usepackage{color}
\usepackage{MnSymbol}

\usepackage{CJK}

\def \be {\begin{equation}}
\def \ee {\end{equation}}
\def \bea {\begin{eqnarray}}
\def \eea {\end{eqnarray}}

\begin{document}

\begin{CJK*}{UTF8}{gbsn}


\title{A topological realization of spin polarization through vortex formation in collisions of Bose-Einstein condensates}



\author{Jian Deng ({\CJKfamily{gbsn}邓建})}
\email{jdeng@sdu.edu.cn}
\affiliation{Institute of Frontier and Interdisciplinary Science,
Key Laboratory of Particle Physics and Particle Irradiation (MOE), Shandong University,
Qingdao, Shandong 266237, China}

\author{Qun Wang ({\CJKfamily{gbsn}王群})}
\email{qunwang@ustc.edu.cn}
\affiliation{Department of Modern Physics, University of Science and Technology of China, Hefei, Anhui 230026, China}

\author{Hong Zhang ({\CJKfamily{gbsn}张宏})}
\email{hong.zhang@sdu.edu.cn}
\affiliation{Institute of Frontier and Interdisciplinary Science,
Key Laboratory of Particle Physics and Particle Irradiation (MOE), Shandong University,
Qingdao, Shandong 266237, China}


\date{\today}

\begin{abstract}
The global spin polarization of hadrons in heavy ion collisions has
been measured in STAR (the Solenoidal Tracker At Relativistic heavy ion collider) experiments, which opens up a new window in the study of the hottest, least viscous and most vortical fluid that
has ever been produced in the laboratory. We present a different approach
to spin polarization from conventional ones: a topological realization
of spin polarization through quantum vortex formation in collisions
of Bose-Einstein condensates (BEC). This approach is based on the observation that the vortex is a topological excitation in a superfluid in presence of local orbital angular momentum and is an analogue of spin degrees of freedom. The formation processes of vortices and vortex-antivortex pairs are investigated by solving the Gross-Pitaevskii Equation with
a large-scale parallel algorithm on Graphics Processing Unit (GPU) to very high precision. In a rotating environment, the primary vortex with winding number one
is stable against perturbation, which has a minimal energy and fixed orbital angular momentum (OAM), but the vortices with larger winding numbers are unstable and
will decay into primary vortices through a redistribution of the energy
and vorticity. The injection of OAM can also be realized in non-central
collisions of self-interacting condensates, part of the OAM of the
initial state will induce the formation of vortices through concentration
of energy and vorticity density around topological defects. Different
from a hydrodynamical description, the interference of the wave function
plays an important role in the transport of energy and vorticity,
reflecting the quantum nature of the vortex formation process. The
study of the vortex formation may shed light on the nature of particle
spin and spin-orbit couplings in strong interaction matter produced
in heavy-ion collisions.
\end{abstract}


\maketitle 
\end{CJK*}

\section{Introduction}

The rotation of a free body induced by a change in its magnetic moment
is called the Einstein-de Haas effect which was discovered in 1915
\cite{dehaas:1915}. It is a consequence of the angular momentum conservation.
A twin effect that the rotation of an uncharged material leads to
a spontaneous magnetization was discovered by Samuel Barnett in the
same year and named the Barnett effect \cite{Barnett:1935}. These
two effects reveal an intrinsic connection between the rotation and
spin polarization in materials.

In ultra-relativistic heavy-ion collisions, similar phenomena also
exist: huge orbital angular momenta (OAM) can be generated in peripheral
collisions with respect to the reaction plane. A part of OAM is transferred
to the hot and dense medium in the form of the global polarization
of hadrons \cite{Liang:2004ph,Liang:2004xn,Voloshin:2004ha,Betz:2007kg,Becattini:2007sr,Gao:2007bc}
(see, e.g. \cite{Wang:2017jpl,Florkowski:2018fap,Gao:2020vbh,Gao:2020lxh,Huang:2020dtn,Becattini:2020ngo,Becattini:2020sww},
for recent reviews). The microscopic mechanism of converting OAM to
hadrons' spin polarization is the spin-orbit coupling in particle
scatterings \cite{Liang:2004ph,Gao:2007bc,Zhang:2019xya,Weickgenannt:2020aaf,Weickgenannt:2021cuo,Sheng:2021kfc},
resulting in the spin-vorticity coupling of the fluid \cite{Becattini:2013fla,Csernai:2013bqa,Becattini:2013vja,Becattini:2016gvu,Fang:2016vpj,Pang:2016igs,Florkowski:2017dyn,Florkowski:2018ahw}
when an ensemble average over randomized collisions is taken. The
global polarization hyperons such as $\Lambda$ (including $\overline{\Lambda}$)
can be measured through their weak decays into hadrons that break
the parity \cite{Liang:2004ph}. A non-vanishing global polarization
of $\Lambda$ hyperons in Au+Au collisions at $\sqrt{s_{NN}}=$ 7.7-200
GeV was measured in STAR experiments \cite{STAR:2017ckg,STAR:2018gyt}.

It is hard to measure the polarization of mesons from their strong
decays due to parity conservation. However, the 00-component of the
spin density matrix $\rho_{00}$ for the vector meson can be measured
through the angular distribution of its decay daughter in its rest
frame. In the unpolarized case, $\rho_{00}$ should be 1/3, which
implies that three spin states in the spin quantization direction
(say, the $z$-direction) with $S_{z}=-1,0,1$ are equally populated.
When $\rho_{00}$ is not 1/3, three spin states are not equally populated
and the spin is called to be aligned in the spin quantization direction.
Recently a large positive deviation of $\rho_{00}$ from 1/3 for $\phi$
mesons has been measured by the STAR collaboration \cite{STAR:2022fan},
which may be explained by fluctuating strong force fields \cite{Sheng:2019kmk,Sheng:2020ghv}.

The global and local polarization in heavy ion collisions are usually
described by hydrodynamics or transport models through the spin-vorticity
couplings \cite{Csernai:2013bqa,Becattini:2013vja,Becattini:2016gvu,Pang:2016igs,Jiang:2016woz,Florkowski:2017dyn,Li:2017slc,Xia:2018tes,Wei:2018zfb,Fu:2021pok,Becattini:2021iol}.
It has been shown that the spin-vorticity couplings can be realized
by non-local collisions of particles which convert the local orbital
angular momentum to the spin polarization \cite{Zhang:2019xya,Weickgenannt:2020aaf,Weickgenannt:2021cuo}.

In this paper, we will present another approach to spin polarization:
a topological realization of spin polarization through quantum vortex
formation in collisions of Bose-Einstein condensates (BEC). Quantum
vortex in BEC originates from topological defects and is an analogue
of spin degrees of freedom. Therefore the formation process of quantum
vorticies in collisions of BEC may shed light on the nature of the
particle spin as well as spin-orbit and spin-vorticity couplings in
the strong-interaction matter.

Although the collisions of BEC in this work will be described by the Gross-Pitaevskii equation for non-relativistic superfluids, they can be generalized to relativistic superfluids in the future. There are a few examples for relativistic superfluids. One example is the relativistic BEC far from equilibrium which can be formed dynamically in an over-populated initial state of gluons at the early stage of heavy ion collisions \cite{Berges:PRL2012,BLAIZOT201268}. Another example is the pion condensation in the cold and dense nuclear matter \cite{BROWN19761}. At zero temperature, the isospin asymmetric quark matter will undergo a phase transition into the pion condensate when the isospin chemical potential is larger than the pion mass \cite{Son:prl2001,Splittorff:prd2001,He:2005nk,Xia:2013caa,Liu:2021gsi}. These relativistic superfluids in heavy ion collisions or neutron stars can be described by relativistic hydrodynamics  \cite{Torres-Rincon:2022ssx}. The infrared dynamics of relativistic scalar fields can be mapped to non-relativistic complex scalar fields, so that the universal scaling of infrared modes and vortex dynamics in superfluids can be studied in both relativistic and non-relativistic regimes \cite{Deng:2018xsk}. In the present work, we will focus on the vortex dynamics of non-relativistic systems. The relativistic version of the same study will be addressed in the future.

The outline of the paper is as follows. In Sec. \ref{vortex-bec}, we will give an introduction to vortex formation in BEC. In Sec. \ref{sec:GPE}, we will derive the Gross-Pitaevskii equation and the conservation laws for the particle number, energy-momentum, OAM and vorticity. In Sec. \ref{sec:vortex1}, we will discuss the stationary properties of a single vortex in the uniform BEC. Comparing to the vortex with winding number one (called primary vortex), the vortex with a larger winding number is unstable and subject to decay into primary vortices. In Set. \ref{sec:vortex2}, with a high precision numerical simulation  performed on Graphics Processing Unit (GPU), we will demonstrate the stability of the primary vortex as well as the decay process of vortices with larger winding numbers. In Set. \ref{sec:collision}, we will perform a simulation of the non-central collision process of two BECs. The finite OAM of the initial state can generate an array of vortices, where the energy and vorticity are concentrated around topological defects. With larger collision velocities, vortex-antivortex pairs can also be generated without initial seeds of topological defects. The conclusions and discussions will be made in Set. \ref{sec:conclusion}.

\textit{Convention}. The coordinate three-vector $\mathbf{r}$ is denoted in boldface and its three components are denoted as $r_{i}$ with $i=1,2,3$ or $x,y,z$ corresponding to three basis vectors $(\mathbf{e}_1,\mathbf{e}_2,\mathbf{e}_3)$ or $(\mathbf{e}_x,\mathbf{e}_y,\mathbf{e}_z)$. The space derivative operator as a three-vector is denoted as $\boldsymbol{\nabla}=(\partial /\partial r_1, \partial /\partial r_2, \partial /\partial r_3)\equiv (\partial_1, \partial_2, \partial_3)$.

\section{A brief introduction to vortex formation in BEC}
\label{vortex-bec}

Quantized vortices are soliton-like excitations in rotating BEC arising from topological defects in the wave function \cite{Fetter:2009zz}. Unlike the whirls in classical hydrodynamics, the vortices in BEC do not change the kinetic energy into the thermal energy. In fact, part of the energy and orbital angular momentum (OAM) are stored in vortices.

Because of the close connection between the vortex and OAM, vortices can be created by an OAM injection into BEC in many different ways. There are vortices in a rotating superfluid Helium-4, so rotation is an effective method to produce quantized vorticies in the superfluid. But this method of normal rotation has difficulties for BEC in a dilute system of cold atoms \cite{Matthews:1999zz}, other methods have been developed other than the rotation \cite{Fetter:2009zz}. As the circulating motion around a vortex is controlled by the winding phase of the wave function, a vortex can be created by manipulating the quantum phase of the condensate, including a dynamical phase-imprinting \cite{Matthews:1999zz}, or a topological phase imprinting \cite{PhysRevLett.89.190403}. Quantum vortices can also be excited by sweeping a laser beam through a condensate at a proper velocity, where the local OAM is injected by the  stirring \cite{Inouye:2001}. Highly ordered vortex lattices can be created in a fast rotating deformed condensate, and the rotation is driven by an optical dipole force \cite{Shaeer:2001}. The stable vortex patterns in a rotating and weakly interacting Bose gas can be well predicted by minimizing the total energy with variational wave functions of the condensate \cite{Butts:1999}. For the rotating and strongly interacting superfluid helium, the vortex lattices can be reproduced by introducing the Magnus force and interaction forces among vortices into the two-fluid hydrodynamics model \cite{Tsuzuki:2021}. Even without additional stirring or rotation, vortices can still be created through merge and interference of multiple BEC clouds, which are initialized to static, independent and uncorrelated states \cite{Scherer:2007}. The relative OAM can be initially stored in relative phases among different condensates and released into the mixture to generate vortices after the removal of the separating optical barrier \cite{Carretero:2008}.

As topological defects, bound vortex-antivortex pairs or unbound vortices can be generated in a two-dimensional system in thermal equilibrium through the Berezinskii-Kosterlitz-Thouless (BKT) topological phase transition \cite{Berezinskii,Kosterlitz:1973xp}. The construction pattern of vortices is employed to define a topological order. This long range order is based on the overall property of the system rather than the behavior of a two-point correlation function, the latter is used to define a traditional phase transition while the former is associated with a topological phase transition.

In a system crossing the second-order phase transition, vortices can be spontaneously created by the growing and merging of domains with independent order parameters, known as Kibble-Zurek mechanism (KZM) \cite{Kibble:1976sj,Zurek:1985qw,Hindmarsh:1994re,Zurek:1996sj}. The KZM is originally introduced in cosmology, but it can be applied to a wide variety of physical systems. Related to such a mechanism, the formation of Kibble-Zurek solitons in a Bose-Einstein condensate has also been extensively studied \cite{Scherer:2007,Carretero:2008,Weiler:2008,Campo:2011,Lamporesi:2013,Chesler:2014gya}.

Vortex-antivortex pairs can also be created in an over-populated boson system through strong turbulence \cite{Deng:2018xsk}. The dynamics of vortex plays an important role in understanding how the off-equilibrium system evolves toward a quasi-stationary nonthermal fixed point \cite{Berges:2008wm,Berges:2008sr,Berges:2014bba}. The creation, scattering and decay of vortex-antivortex pairs are strongly correlated with the scaling properties of the self-similar evolution \cite{Nowak:2011sk,Schole:2012kt,Karl:2016wko}.
The strongly anomalous scaling can be understood through three-body collisions of vortex-antivortex pairs with unbound vortices as well as collisions between vortex-antivortex pairs \cite{Deng:2018xsk, Karl:2016wko}.

Vortices can be formed in cold dark matter halos if they have sufficient number density, angular momentum and self interaction \cite{Rindler-Daller:2011afd}. When de Broglie wavelengths of bosonic dark matter particles are larger than average inter-particle distances, the interference of matter wave inside a halo can also leads to vortex string or vortex rings with non-trivial windings \cite{Hui:2020hbq}. The formation, deformation and reconnection of vortex rings in BECs or in viscous fluids are intensively studied\cite{Nguyen:2019, zou:2021, Nguyen:2021}. It would be interesting to study the signature of these vortex structures in the dark matter superfluid, for instance, through gravitational lensing \cite{Berezhiani:2015pia,Berezhiani:2015bqa}. If low-density vortices in galaxies can be confirmed, the superfluidity can potentially distinguish the scenarios of some dark matter models.

During the vortex nucleation in a rotating BEC, the symmetry breaking and quantum nature are studied by comparing the symmetry properties of the true many-body state and its mean-field approximation \cite{Dagnino:2009}. At the critical frequency, the exact many-body ground state exhibits strong correlation and entanglement, while in the mean-field approximation there is a jump in the order parameter reflecting the change of symmetry. Far from the critical frequency, the mean-field theory can well describe the formation process of vortices.

\section{Gross-Pitaevskii equation and conservation laws}
\label{sec:GPE}

For the quantum state of a boson gas, let us define $\Psi^{\dagger}(\mathbf{r})$ and $\Psi(\mathbf{r})$ as the field operators to create and annihilate a particle at point $\mathbf{r}$ respectively. These operators satisfy the bosonic commutation relation,
\bea
\left[\Psi(\mathbf{r}), \Psi^{\dagger}(\mathbf{r}')\right] &=& \delta(\mathbf{r}-\mathbf{r}'), \nonumber \\
\left[\Psi(\mathbf{r}),\Psi(\mathbf{r}')\right] &=& \left[\Psi^{\dagger}(\mathbf{r}), \Psi^{\dagger}(\mathbf{r}')\right] = 0.
\eea
The ensemble average $\langle \Psi^{\dagger}(\mathbf{r})\Psi(\mathbf{r}) \rangle$ gives the expectation value of the number density $n(\mathbf{r})$ with the total number given by $N=\int d\mathbf{r} n(\mathbf{r})$. If the interacting range $r_0$ and the $s$-wave scattering length $a_s$ are much smaller than the average distance between particles $d=(N/V)^{-1/3}$, the binary interaction will dominate the dilute and weakly interacting gas. The dynamics of the system is controlled by the Hamiltonian operator,
\bea
\hat{H}&=& \int \frac{\hbar^2}{2m} \boldsymbol{\nabla} \Psi ^{\dagger}(\mathbf{r},t) \cdot \boldsymbol{\nabla} \Psi (\mathbf{r},t)  d^3\mathbf{r} + \int  \frac{1}{2} \Psi ^{\dagger}(\mathbf{r},t) \Psi ^{\dagger}(\mathbf{r}',t)V(\mathbf{r}-\mathbf{r'}) \Psi (\mathbf{r},t) \Psi (\mathbf{r}',t) d^3\mathbf{r}d^3\mathbf{r}',
\eea
where $\hbar$ is the reduced Planck constant, $m$ is the particle mass and $V(\mathbf{r}-\mathbf{r'})$ is the two-body potential. In the zero temperature limit, almost all particles are in  the ground state to form the Bose-Einstein condensation (BEC). Because the total number is huge, the non-commutability of the field operators can be ignored. In this situation, we can replace the field operator $\Psi (\mathbf{r},t)$ with a classical field $\psi(\mathbf{r},t)$ so the Hamiltonian operator $\hat{H}$ becomes the classical Hamiltonian $H$. The classical field $\psi(\mathbf{r},t)$ is also called the wave function of the BEC but there is a difference from the normal wave function in the Schr\"odinger equation or quantum mechanics for single particles in the normalization: in the BEC case, $\psi(\mathbf{r},t)$ is normalized to the particle number $N$, while in the latter case, the wave function is normalized to 1. In the rest of the paper, $\psi(\mathbf{r},t)$ will be called either the classical field or the wave function: both are equivalent.

For the behavior of the classical field over a longer distance than the scattering length, the details of the potential are not relevant any more. We can approximate the potential by a contact form $V(\mathbf{r}-\mathbf{r'})\approx \lambda \delta^{3}(\mathbf{r}-\mathbf{r'})$, as long as the Born approximation is applicable and the coupling constant $\lambda$ can reproduce the $s$-wave scattering length, $\lambda= 4\pi a_s \hbar^2/m$. So the Hamiltonian for a BEC at zero temperature can be written as,
\bea
H=\int  \mathcal{H} d^3\mathbf{r} = \int \left[\frac{\hbar^2}{2m} \boldsymbol{\nabla} \psi(\mathbf{r},t) \cdot \boldsymbol{\nabla} \psi^*(\mathbf{r},t) + \frac{\lambda}{2} |\psi(\mathbf{r},t)|^4 \right]d^3\mathbf{r} , \label{Hamiltonian}
\eea
The classical field $\psi(\mathbf{r},t)$ varies slowly in time and space and satisfies a non-linear Schr\"odinger equation, the so called Gross-Pitaevskii Equation (GPE)
\bea
i\hbar \frac{\partial}{\partial t} \psi(\mathbf{r},t) = \frac{\delta H}{\delta \psi^*(\mathbf{r},t)}=-\frac{\hbar^2 }{2m}  \boldsymbol{\nabla}^2 \psi(\mathbf{r},t) + \lambda |\psi(\mathbf{r},t)|^2\psi(\mathbf{r},t).
\label{GPE}
\eea
From this equation, it is easy to obtain conservation laws about the particle number, energy and momentum
\bea
\frac{\partial }{\partial t}\rho  + \boldsymbol{\nabla} \cdot \boldsymbol{j} =0,                  \label{N-con} \\
\frac{\partial }{\partial t}\mathcal{H}  + \boldsymbol{\nabla} \cdot \vec{\mathcal{P}} =0, \label{E-con} \\
\frac{\partial }{\partial t}m\boldsymbol{j}_i + \partial_k \Pi_{ki} =0,              \label{P-con}
\eea
where $\rho$, $\boldsymbol{j}$, $\mathcal{H}$, $ \vec{\mathcal{P}}$, $m\boldsymbol{j}$ and $\Pi$  are the particle number density, the particle number current density, the energy density, the energy current density, the momentum density and the momentum current density, respectively. The classical field can be parameterized in terms of the number density $\rho = |\psi|^2$ and the complex phase angle $\theta(\mathbf{r},t)$,
\begin{equation}
\psi(\mathbf{r},t)=\sqrt{\rho(\mathbf{r},t)}e^{i\theta(\mathbf{r},t)}\,,
\end{equation}
which is called the Madelung representation.  In the Madelung representation the quantities in conservation laws can be put into a compact form with the help of two vectors,
\bea
\mathbf{v}& =&\frac{\hbar}{\rho m }\text{Im}(\psi^* \boldsymbol{\nabla} \psi)=\frac{\hbar}{m}\boldsymbol{\nabla} \theta,  \label{v-def} \\
\mathbf{K}& =& \frac{\hbar}{\rho m }\text{Re}(\psi^* \boldsymbol{\nabla} \psi)=\frac{\hbar}{m}\boldsymbol{\nabla} \ln\sqrt{\rho}. \label{K-def}
\eea
In the particle number conservation law (\ref{N-con}), the particle current is defined as $\boldsymbol{j}= \rho \mathbf{v}$. In the energy conservation law (\ref{E-con}), the energy density has the form,
\bea
\mathcal{H}=\frac12 m \rho (|\mathbf{v}|^2 + |\mathbf{K}|^2) + \frac{\lambda}{2}\rho^2,
\label{En-dens}
\eea
which is composed of a "classical" kinetic part, a "quantum-pressure" part and an interacting part \cite{Nowak:2010tm}. Correspondingly, the energy current density is given by,
\bea
\vec{\mathcal{P}}= \mathcal{H}\mathbf{v} + m \rho (\mathbf{v}\times \mathbf{K}) \times \mathbf{K} + \frac{\rho}{2}\left[(\boldsymbol{\nabla} \cdot \mathbf{v})\mathbf{K} - (\boldsymbol{\nabla} \cdot \mathbf{K}) \mathbf{v}+\lambda \rho \mathbf{v}\right]\,. \label{P-dens}
\eea
In the momentum conservation law (\ref{P-con}), the momentum density is given by $m\boldsymbol{j}$, while the momentum tensor density is given by,
\bea
\Pi_{ij} = m\rho K_{i}K_{j}+m\rho v_{i}v_{j}+\delta_{ij}\frac{\lambda}{2}\rho^{2}-\delta_{ij}\frac{\hbar}{2}\boldsymbol{\nabla}\cdot(\rho\mathbf{K}) \,,
\label{Pi-dens}
\eea
which is symmetric for interchange of two indices. With these ingredients defined, Eq.(\ref{N-con},\ref{E-con},\ref{P-con}) make up a quantum hydrodynamic description for BECs with spin{\textendash}0 Bosons. For BECs with spin{\textendash}1 bosons, a quantum hydrodynamic model is derived in \cite{Andreev:2021}.

For quantum hydrodynamics, the variation in density is not negligible if the "quantum pressure" is comparable to the interaction energy density. The balance between them determines the healing length $\xi = \hbar/\sqrt{2m\lambda\rho}$, over which a perturbation in density can relax to its bulk value. When $\xi \ll D$ with $D$ being the characteristic length scale of density variation, the Euler's equations for the irrotational compressible perfect fluid can be reproduced by just setting $\mathbf{K}=0$ in Eqs.(\ref{En-dens},\ref{P-dens},\ref{Pi-dens}).
From the Euler's equations, we know that the interaction energy density also works as the classical pressure $p=(\lambda/2)\rho^2$. The speed of sound in BEC is determined by the compressibility of the fluid,
\bea
c_s^2 = \frac1m \frac{\partial p}{\partial \rho} = \frac{\lambda \rho}{m}, ~~\text{or}  ~~ c_s =\frac{\hbar}{\sqrt{2}m\xi}.
\eea
The fluid can be considered incompressible if the velocity is small in comparison with $c_s$.
In the matter wave picture, the flow velocity is proportional to the gradient of the wave function's phase $\theta$ which changes $2\pi$ over de Broglie wave length $\lambda_D$, so we have $|\mathbf{v}| \sim (\hbar/m) 2\pi/\lambda_D$. If the flow velocity is comparable with the speed of sound, the particle number density may change dramatically, so that the destructive interference of matter waves are observable at the typical length scale $\lambda_D \sim 2\sqrt{2}\pi \xi$.

It is interesting to make comparison of these length scales,
\bea
\frac{\xi}{d} = \frac{1}{\sqrt{8\pi}}\frac{1}{(\rho a_s^3)^{1/6}} = \sqrt{\frac{d}{8\pi a_s}}.
\eea
Therefore the condition $\lambda_D \gg \xi \gg d \gg a_s$ holds for the dilute Boson gas in the weakly interacting limit.

\subsection{Conservation of OAM and Vorticity}

The Lagrangian of the system can be derived by a Legendre transformation from the Hamiltonian (\ref{Hamiltonian}),
\bea
\mathcal{L} = \frac{i}{2}\left[\psi^* \left(\frac{\partial}{\partial t}{\psi}\right) - \psi \left(\frac{\partial}{\partial t}{\psi}^*\right)\right]- \frac{1}{2m} \boldsymbol{\nabla} \psi \cdot \boldsymbol{\nabla} \psi^* - \frac{\lambda}{2} |\psi|^4.  \label{Lag-den}
\eea
The GPE (\ref{GPE}) is the result of the Euler-Lagrangian equation. The Lagrangian has a global U(1) symmetry and is invariant under translation of time-space and space rotation, leading to the conservation of particle number, energy-momentum and OAM, respectively.

The conservation equation for the OAM reads,
\begin{equation}
\frac{\partial}{\partial t}M_{0j}+\partial_i M_{ij}=0,
\end{equation}
where $M_{0i}$ and $M_{ij}$ ($i,j=1,2,3$) are the OAM density and the $i$-th component of the OAM current density in the $j$ direction, respectively. They are given by,
\begin{eqnarray}
M_{0i} & = & (\mathbf{r}\times m \rho\mathbf{v})_{i}= m (\mathbf{r}\times\boldsymbol{j})_{i},\nonumber \\
M_{ij} & = & \epsilon_{jkl}r_{k}\Pi_{il}\,,
\label{oam-conservation}
\end{eqnarray}
where $\boldsymbol{j}= \rho \mathbf{v}$ is the current density of the particle number and $\Pi_{il}$ is the momentum tensor density given by (\ref{Pi-dens}). The OAM density vector is then $\boldsymbol{l}=m (\mathbf{r}\times\boldsymbol{j})$.

We can also rewrite the OAM conservation equation (\ref{oam-conservation}) into the conservation equation of the vorticity,
\begin{equation}
\frac{\partial}{\partial t} \mathcal{M}_{0j} + \partial_i \mathcal{M}_{ij} =0,
\label{O-con}
\end{equation}
where the vorticity density $\mathcal{M}_{0j}$ and vorticity current density $\mathcal{M}_{ij}$ in the $j$ direction can be obtained by replacing $\mathbf{r}\rightarrow (1/m)\boldsymbol{\nabla}$ and  $r_{k}\rightarrow (1/m)\partial_k$ from $M_{0j}$ and $M_{ij}$ in Eq. (\ref{oam-conservation}), respectively. They are given by,
\begin{eqnarray}
\mathcal{M}_{0j}&=& \left(\boldsymbol{\nabla} \times \boldsymbol{j}\right)_j\,,\\
\mathcal{M}_{ij} &=& \frac{1}{m}\epsilon_{jkl} \partial_k \Pi _{il}\,.
\end{eqnarray}
Eq. (\ref{O-con}) can also be derived by taking a curl of  Eq. (\ref{P-con}) or directly from the GPE. The curl of the momentum current stands for the local density of angular momentum or the vorticity density, which leads to conservation of the total vorticity $\int d^3\mathbf{r} \boldsymbol{\nabla} \times \boldsymbol{j} $. We can rewrite,
\begin{equation}
\boldsymbol{\nabla} \times \boldsymbol{j} = \frac{i\hbar}{m}\boldsymbol{\nabla} \psi \times \boldsymbol{\nabla} \psi^*
= \frac{\hbar}{m} \boldsymbol{\nabla}\rho\times\boldsymbol{\nabla}\theta
= \frac{2m}{\hbar}\rho\mathbf{K}\times \mathbf{v}\,,
\end{equation}
then Eq. (\ref{O-con}) can be put into a different form,
\bea
\frac{\partial}{\partial t} \left[i \boldsymbol{\nabla} \psi \times \boldsymbol{\nabla}\psi^* \right] - \frac{\hbar}{2m} \partial_j\left[\boldsymbol{\nabla} \psi \times \boldsymbol{\nabla} (\partial_j \psi^*)  + \boldsymbol{\nabla} \psi^* \times \boldsymbol{\nabla} (\partial_j \psi)\right] =0,   \label{O-con-1}
\eea
or in the Madelung representation,
\bea
\frac{\partial }{\partial t} \left(\rho \mathbf{K} \times \mathbf{v}\right) + \partial_j \left[ v_j\left(\rho \mathbf{K} \times \mathbf{v}\right)\right] + \frac{\hbar}{2m} \partial_j \left[\rho (\partial_j \mathbf{K})\times \mathbf{K} + \rho (\partial_j \mathbf{v})\times \mathbf{v}
\right] =0.
\label{V-con}
\eea
If the interaction is not in a contact form, there will be an additional term from the interaction $\int V(\mathbf{r}-\mathbf{r}')\mathbf{D}(\mathbf{r}')\times \mathbf{D}(\mathbf{r})d\mathbf{r}'$, with $\mathbf{D}(\mathbf{r})=\boldsymbol{\nabla} \rho(\mathbf{r})$ be the divergence of the number density.

\section{Stationary vortex solution to GPE} \label{sec:vortex1}

For a rotating BEC, there may exist vortex solutions to the GPE. The classical field has a rotation symmetry alone the vortex string, and can be parameterized in the following form,
\bea
\psi(r,\phi,t) = \sqrt{\rho(r)} e^{i s \phi} e^{-i{\mu t}/{\hbar}},
\label{psi-1}
\eea
where $(r, \phi)$ are polar coordinates in the transverse plane perpendicular to the vortex string, with $r$ being the distance of the field point from the vortex string and $\phi$ being the azimuthal angle, $s$ is the winding number which must be an integer to guarantee the single-value wave function, it is required that $\rho=0$ at $r=0$, and the time dependence of the wave function is fixed by the chemical potential $\mu=\delta H/\delta N$. 

From the field form Eq. \eqref{psi-1}, we can read out the velocity $\mathbf{v} = (\hbar s)/(mr)\hat{\phi} $, and then we obtain the OAM density $\boldsymbol{l} = \rho(r) s\hbar \hat{z}$, which indicates that each particle contributes $s\hbar$ to the total OAM, no matter how far it is from the vortex string. The number density varies strongly in the range of the healing length $\xi$ near the vortex string. For convenience we use the profile function $f(\eta) = \sqrt{\rho(\eta)/\rho_0}$ for the field amplitude where $\rho_0$ is the density in the bulk and $\eta\equiv r/\xi$ is the rescaled radius. Inserting Eq. (\ref{psi-1}) into Eq. (\ref{GPE}), we obtain the equation,
\begin{equation}
\frac{1}{\eta} \frac{d}{d\eta} \left(\eta\frac{d f(\eta)}{d\eta} \right) + \left( \frac{\mu}{\lambda \rho_0}-\frac{s^2}{\eta^2}\right) f(\eta) - [f(\eta)]^3 =0,  \label{Eq:vortex-profile}
\end{equation}
with boundary conditions $f(\eta \rightarrow 0)=0$ and $f(\eta\rightarrow\infty)=1$. To satisfy  boundary conditions, it is easy to see $\mu=\lambda \rho_0$, then the profile function can be numerically solved. A shooting method is employed to solve this non-linear differential equation. By tuning the first order derivative $f^\prime (\eta)$ at $\eta=0$, we can find the solution which satisfies the condition $f(\eta\rightarrow \infty)=1$. We have checked the result with a variational method and a semi-analytical method, all of them can precisely reproduce the profile function for $s =1,2,3$.

\begin{figure}[ht] 
\centering
\includegraphics[scale=0.55]{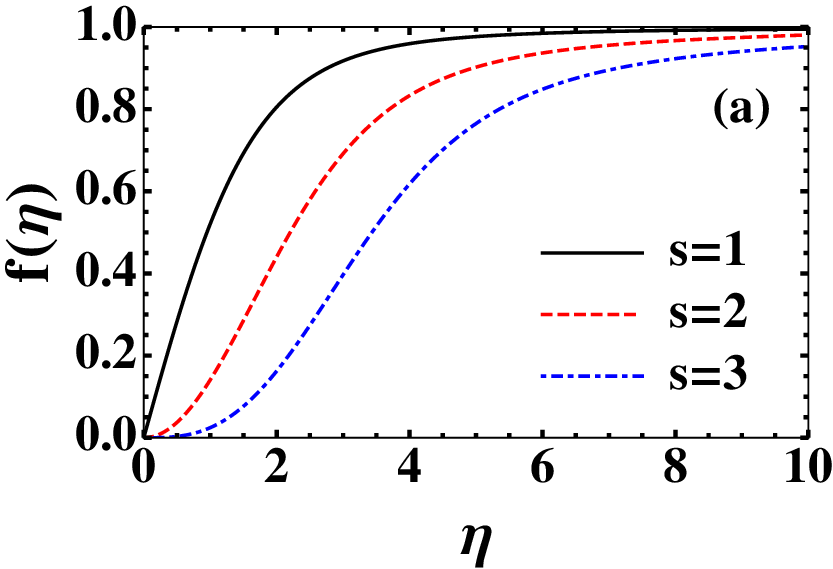}
\includegraphics[scale=0.55]{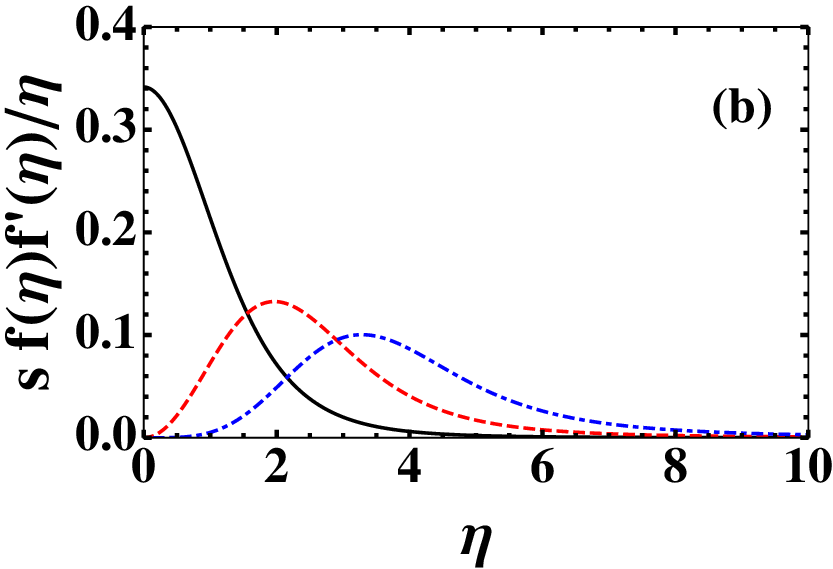}
\includegraphics[scale=0.55]{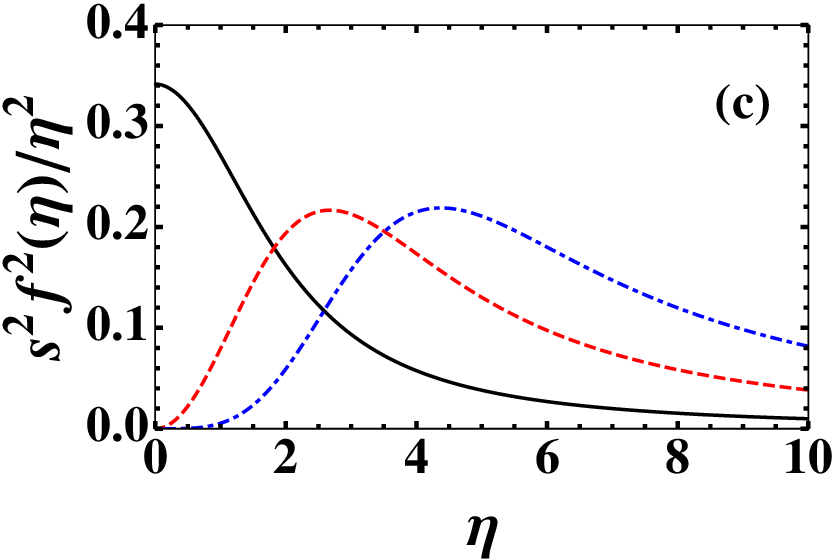}
\includegraphics[scale=0.55]{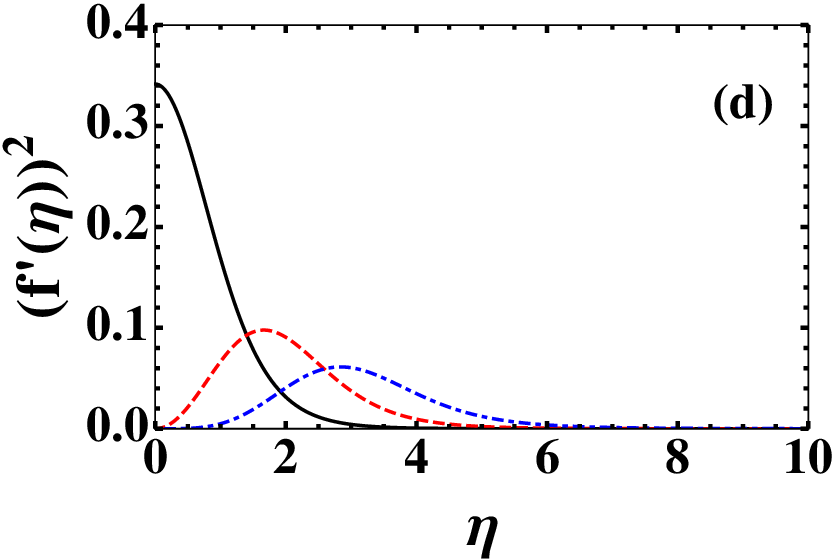}
\caption{Stationary vortices with winding number $s=1,2,3$ are shown in black solid, red dashed and blue dash-dotted lines respectively. (a) The profile of the field $f(\eta)$. (b) The rescaled vorticity density $|\boldsymbol{\nabla}\times \boldsymbol{j}|$. (c) The rescaled classical kinetic energy density $(1/2)m\rho |\mathbf{v}|^2$. (d) The rescaled energy density connected with the quantum pressure $(1/2)m\rho |\mathbf{K}|^2$.
}\label{fig:stationary}
\end{figure}

In Fig. \ref{fig:stationary}, we show the profiles of the field, vorticity density and energy densities as functions of $\eta$. As expected, the field profile around a vortex varies in a distance of a few times the healing length. The field profile increases near the center from zero with a slope $f'(\eta)\propto \eta^s$ and finally approaches unity in the form $f(\eta\rightarrow\infty) \sim \sqrt{1-s^2/\eta^2}$. The field profiles for different winding numbers have similar shapes, but other quantities show qualitative difference between the results of the primary vortex with $s=1$ and other vortices with $s>1$. These quantities are the magnitude of the vorticity density $|\boldsymbol{\nabla}\times\boldsymbol{j}|$ in Fig. \ref{fig:stationary}b, the classical kinetic energy density $(1/2)m\rho|\mathbf{v}|^{2}$ in Fig. \ref{fig:stationary}c and the energy density connected with the quantum pressure $(1/2)m\rho|\mathbf{K}|^{2}$ in Fig. \ref{fig:stationary}d,
\begin{eqnarray}
|\boldsymbol{\nabla}\times\boldsymbol{j}| & = & s\hbar\frac{2\rho_{0}}{m\xi^{2}}\frac{f(\eta)f'(\eta)}{\eta},\nonumber \\
\frac{1}{2}m\rho|\mathbf{v}|^{2} & = & s^{2}\hbar^{2}\frac{\rho_{0}}{2m\xi^{2}}\frac{f^{2}(\eta)}{\eta^{2}},\nonumber \\
\frac{1}{2}m\rho|\mathbf{K}|^{2} & = & \hbar^{2}\frac{\rho_{0}}{2m\xi^{2}}[f'(\eta)]^{2}.
\end{eqnarray}
For the primary vortex, these quantities show a core structure in the profiles, in other words, they have maxima at the vortex center. While for vortices with $s>1$, the profiles of these quantities show a ring structure, which vanish at the vortex center and reach maxima at finite radii. These differences indicate that the primary vortex is topologically different from vortices with larger winding numbers. The primary vortex looks like a particle in the ground state in a rotational BEC, while the vortices with larger winding numbers look like particles in exciting states and will decay into the primary one when they are perturbed by fluctuations \cite{BOOK:2003}. This picture can be better illuminated by an analysis of the energy and OAM carried by vortices \cite{BOOK:2003}.

In grand canonical ensemble, the relative energy of a vortex in the BEC in
a cylinder with the volume $V=\pi L R^2$ is given by,
\bea
\Delta E_s &=& \int_V d^3\mathbf{r} \left[\mathcal{H}(\mathbf{r}) -\mu \rho(\mathbf{r})\right] - V\left(\frac{\lambda }{2} \rho_0^2 - \mu \rho_0 \right )
= \frac{L\pi \rho_0}{m} \int_0^{R/\xi} \eta d\eta \left[\left(\frac{df}{d\eta}\right)^2 + \frac{s^2}{\eta^2} f^2 + \frac{1}{2}\left(f^2-1\right)^2\right] \nonumber \\
\eea
where $\mathcal{H}(\mathbf{r})$ is given in Eq. (\ref{En-dens}). There is a logarithmic divergence from the kinetic energy, so we introduce a radius cutoff $R\gg\xi$ to regulate the integral. With the profile solution $f(\eta)$, the result takes a compact form,
\begin{equation}
\Delta E_s = \frac{L\pi \rho_0}{m} s^2 \ln \frac{ \gamma_sR}{\xi}\,,
\label{vortex-energy}
\end{equation}
where the factor $\gamma_{s} =1.464, 0.572, 0.358$ for $s=1,2,3$ respectively. For each individual vortex, the relative energy is proportional to $s^2$, while the OAM is proportional to $s$, so vortices with larger winding numbers are not stable and will decay to primary vortices. Even the primary vortex is an excited state of BEC because of its positive energy. Only in a rotating frame with a considerable larger angular velocity $\boldsymbol{\omega}$ or in a system with a sizeable OAM can the vortex solution correspond to the global or local minimum of the free energy defined as
$F = \int_V  d^3\mathbf{r} \left[\mathcal{H}(\mathbf{r}) -\mu \rho(\mathbf{r})-\boldsymbol{\omega}\cdot \boldsymbol{l}\right] $.

\section{Evolution of single vortex}
\label{sec:vortex2}

\begin{figure}[ht]
\centering
\includegraphics[scale=0.75]{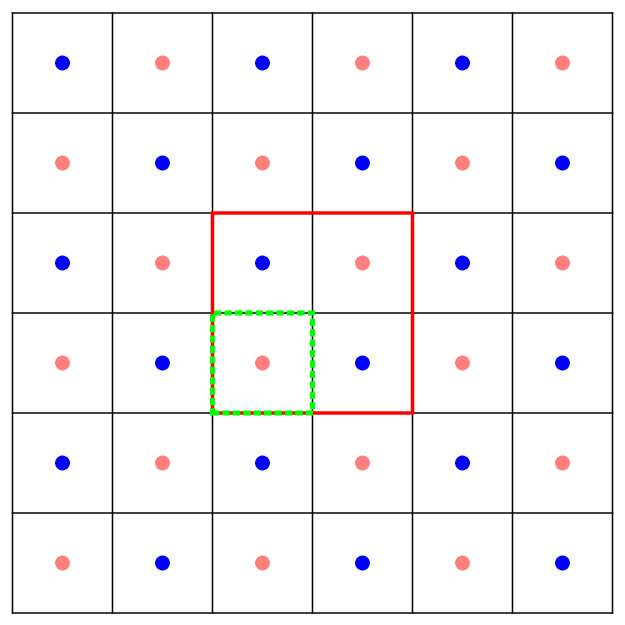}
\caption{
A sketchy illustration of the field configuration at the initial time with vortex and anti-vortex pairs to fulfill periodic boundary conditions. The pink and blue points located at centers of their blocks stand for vortices with positive or negative winding numbers respectively. The red solid lines indicate the boundary of simulation area, in which the wave function is iterated. }
\label{fig:grid}
\end{figure}

To know how vortices evolve with time, we need to solve the GPE (\ref{GPE}) with proper initial conditions and period boundary condition \cite{Koplik:1993}. We choose the following initial value for the classical field,
\begin{equation}
\psi(t=0,\mathbf{r}) = \sqrt{\rho_0} \prod_{n=-N_0}^{N_0-1} \prod_{k=-N_0}^{N_0-1}g\left(\frac{|\mathbf{r}-\mathbf{r}_{nk}|}{\xi}\right)\exp\left[i s_{nk}\phi(\mathbf{r}-\mathbf{r}_{nk})\right]\,,
\label{initial-vortex-array}
\end{equation}
which represents an array of vortices or rotating fields as shown in Fig. \ref{fig:grid} for $N_0=3$. In Eq. (\ref{initial-vortex-array}), $g(\eta)$ stands for an amplitude function satisfying $g(0)=0$ and $g(\eta\rightarrow\infty) \rightarrow 1$, $\phi(\mathbf{r})$ is the polar angle of $\mathbf{r}$, and $2N_0$ is the number of blocks in one dimension. Each vortex is localized at the center position of each block $\mathbf{r}_{nk} = (n+1/2) A \mathbf{e}_x + (k+1/2) A \mathbf{e}_y $ ($A$ is the side length of the block) and carries a local OAM with respect to the center with the winding number $s_{nk}=|s|(-1)^{n+k}$. The winding numbers of neighboring vortices are opposite in sign but have the same absolute value which are represented by different colors in the figure. From the symmetry of the configuration, it can be easily checked that the periodic condition is satisfied by Eq. \eqref{initial-vortex-array} at boundaries $x=\pm A$ and $y=\pm A$. The gradient of the field does not have such a periodic feature for finite $N_0$, but the difference between gradients at two corresponding points on the boundary can be suppressed significantly by increasing $N_0$. With $N_0 \ge 3$, the difference in gradients is negligible.

In the numerical simulation, we set the parameters as $\hbar =1$, $m=4$, $\lambda =0.002$, and $\rho_0 =1$. With these values of parameters the healing length becomes $\xi = \hbar/\sqrt{2m\lambda\rho_0} \sim 7.9$ and the chemical potential becomes $\mu =\lambda\rho_0 = 0.002$ which set the scales in space and time respectively. We employ a symplectic algorithm at sixth order \cite{Yoshida:1990zz, Levkov:2018kau} to perform the time evolution on GPU, and find that the simulation with grid size $a=1$ and time step $\Delta t =0.5$ can maintain the energy conservation and time reversibility in a high precision. A brief introduction to symplectic integration algorithm is given in Appendix \ref{symplectic-integrator}.

\begin{figure}[ht] 
\centering
\includegraphics[scale=0.25]{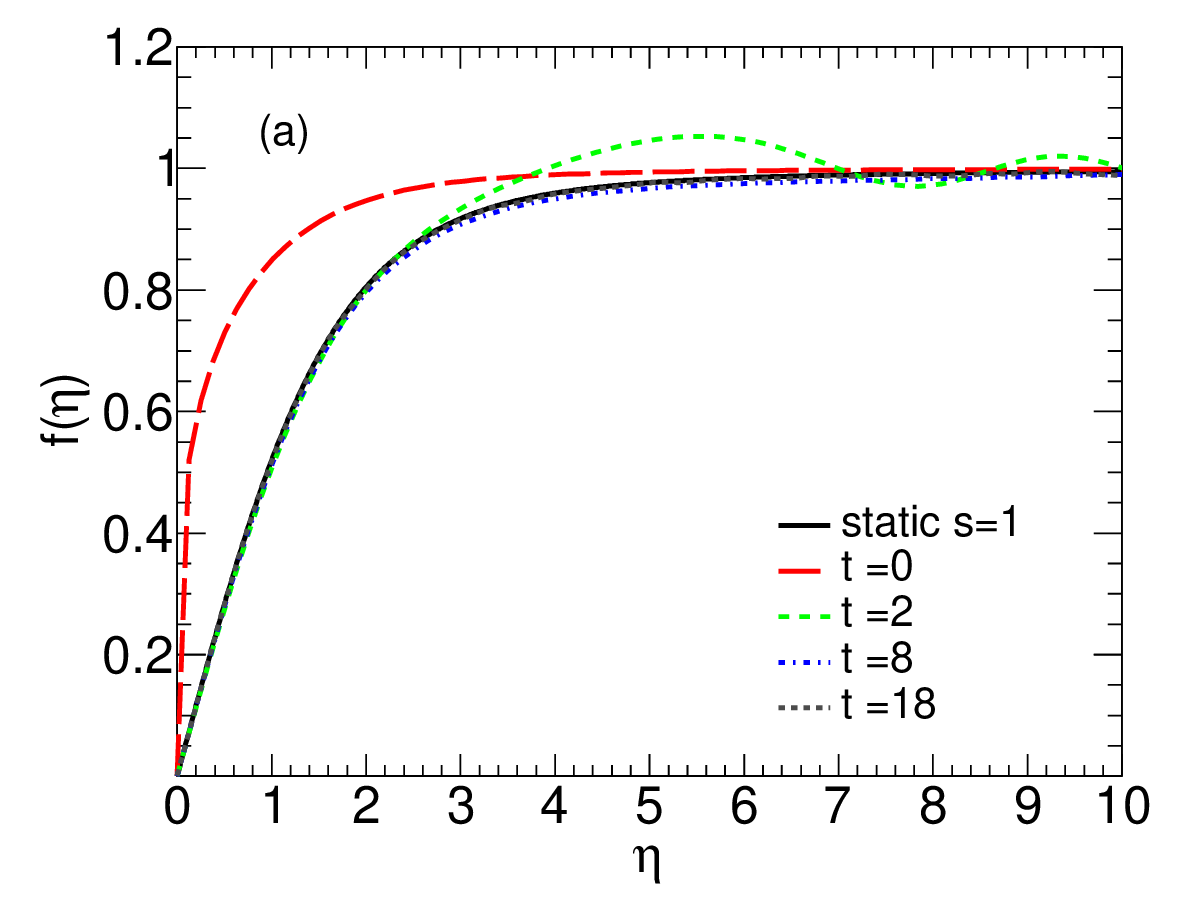}
\includegraphics[scale=0.25]{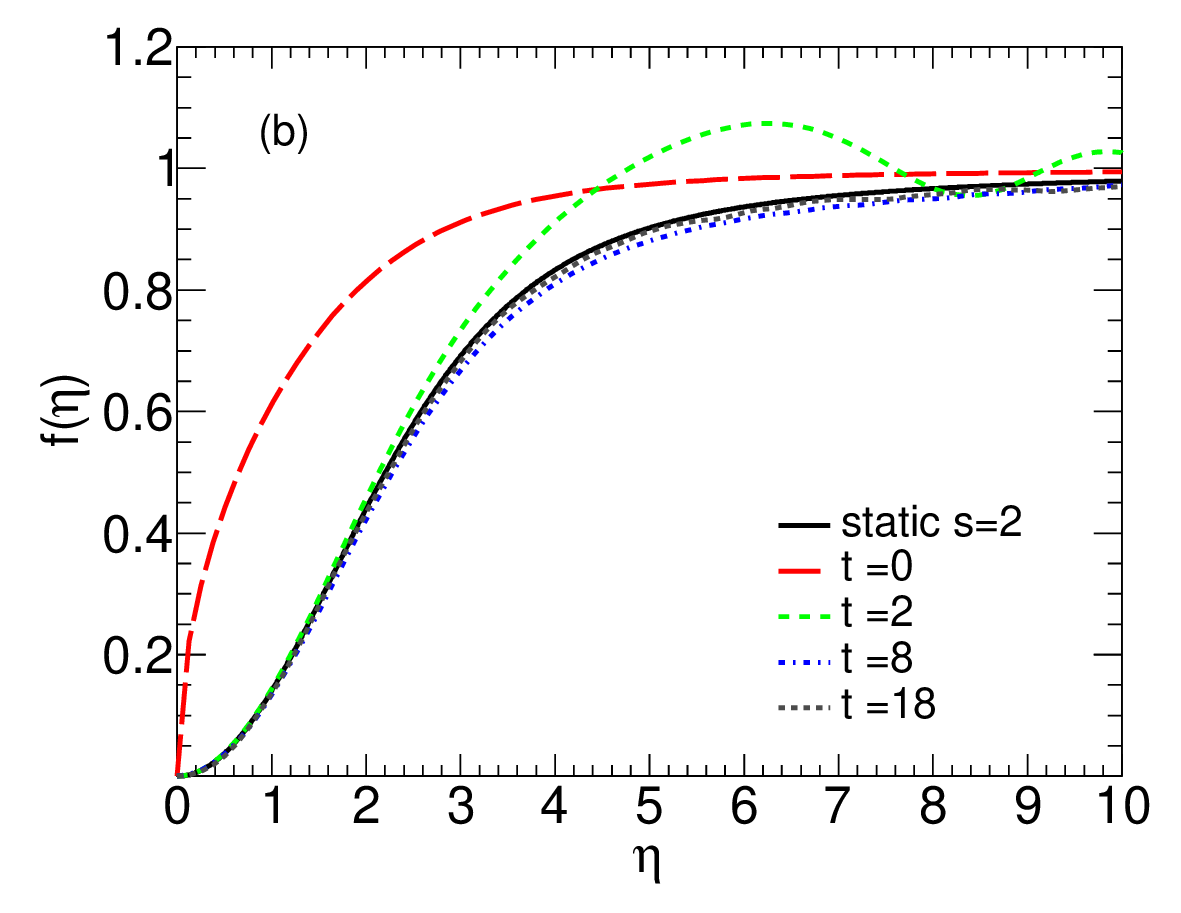}
\includegraphics[scale=0.25]{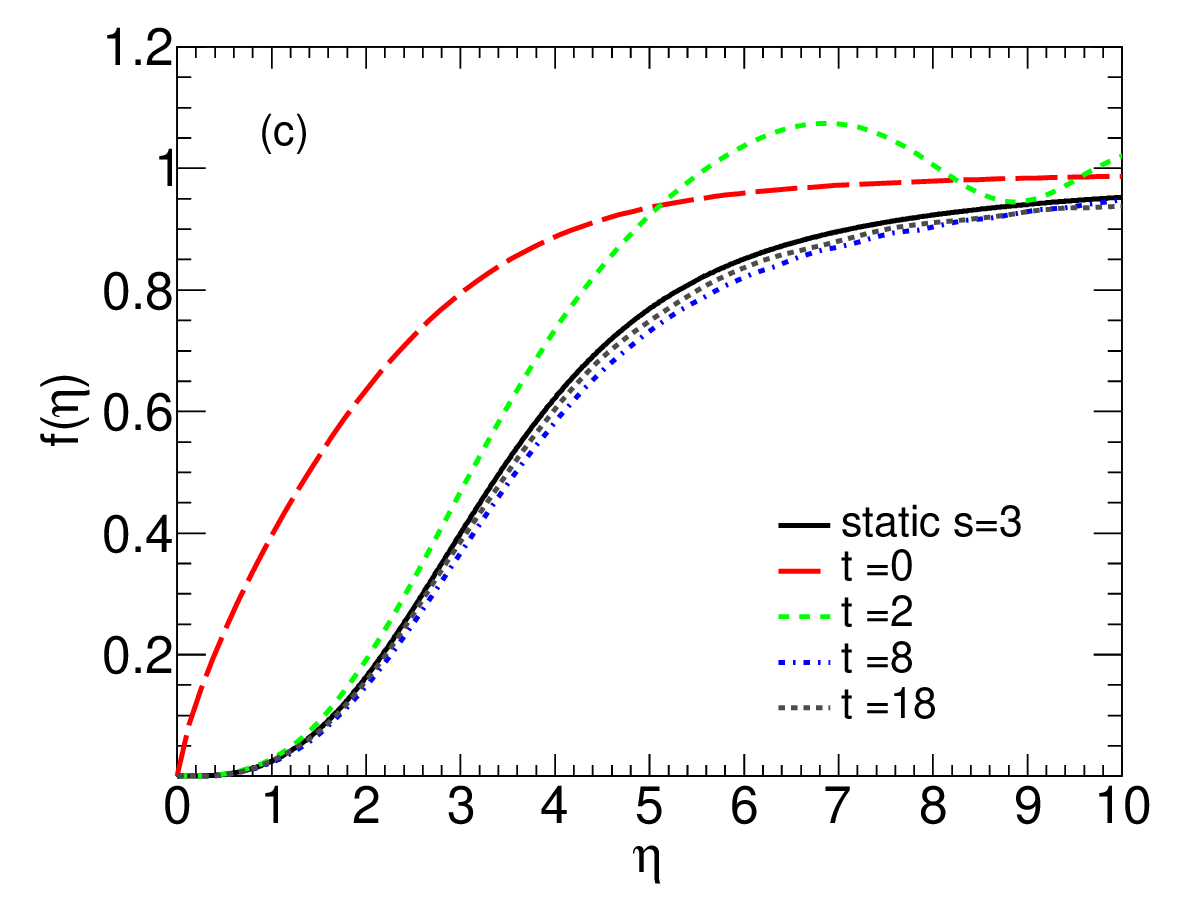}
\includegraphics[scale=0.25]{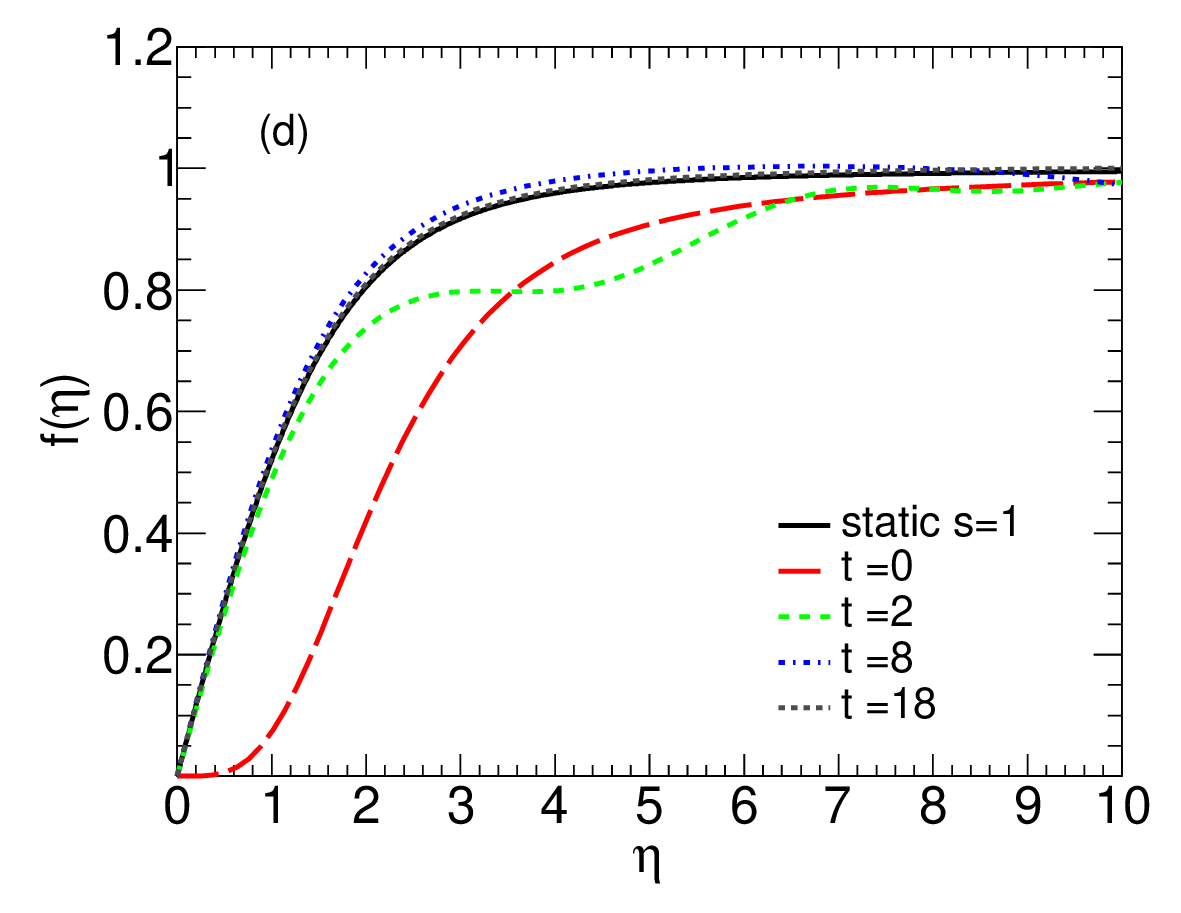}
\includegraphics[scale=0.25]{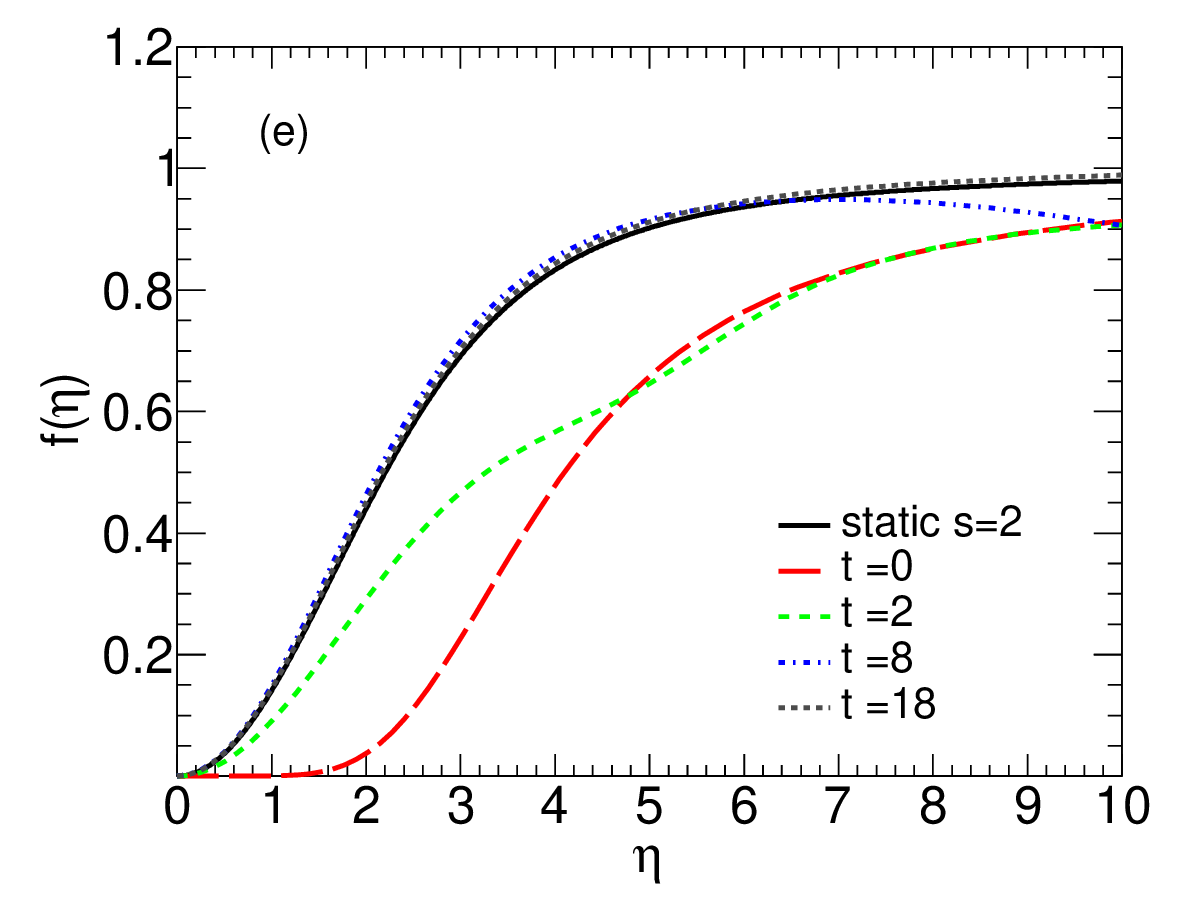}
\includegraphics[scale=0.25]{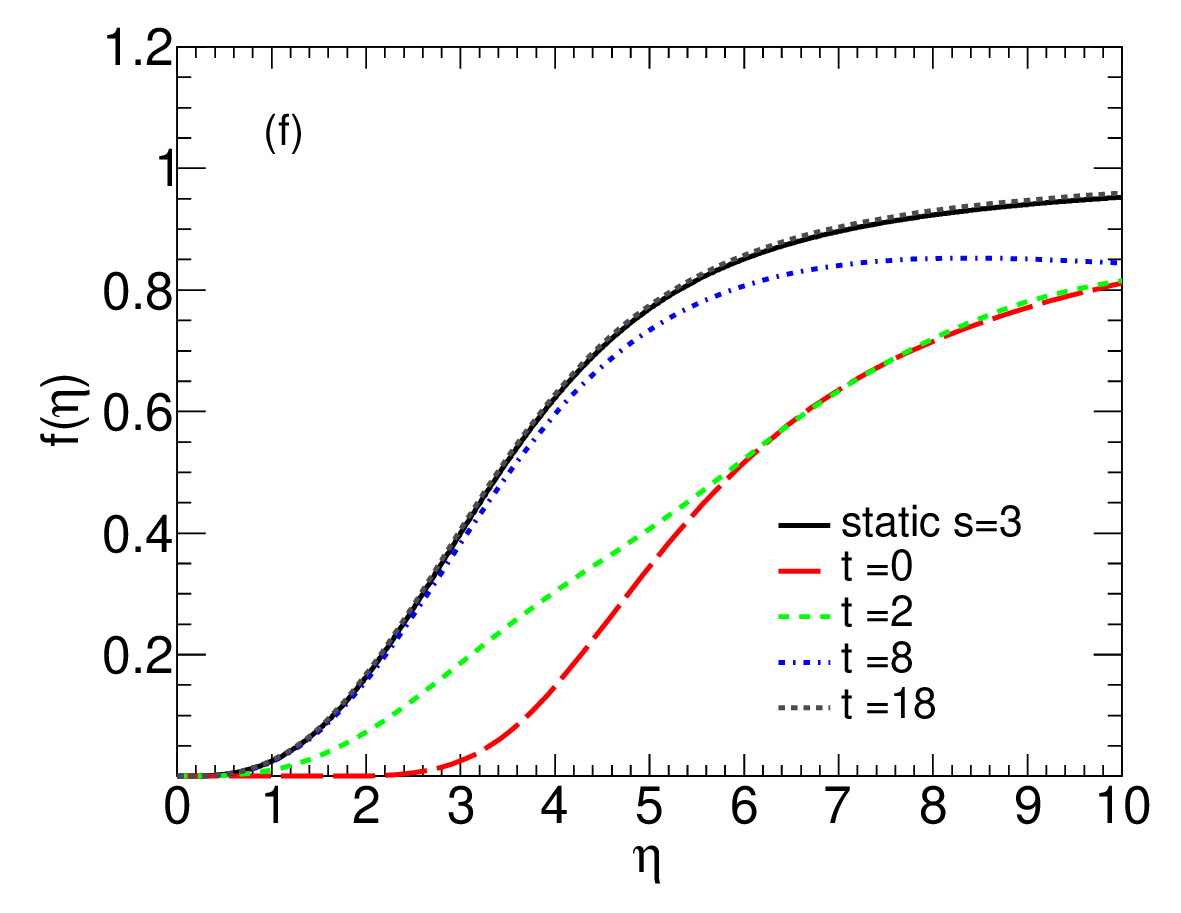}
\caption{Snapshots of density profiles in the early evolution of a rotating BEC. The upper panels (a,b,c) show the results from an initial condition with a larger density, and the lower panels (d,e,f) show the results with a lower density. The left panels (a,d), middle panels (b,e) and right panels (c,f) correspond to the results with a winding number $s = 1, 2, 3$  respectively. The initial profiles are displayed in red-long-dashed lines, while static limits are displayed in black-solid lines. The green-short-dashed, blue-dash-dotted and black-dotted lines are profiles at some intermediate time on the way to static limits.}
\label{fig:goto-static}
\end{figure}

To look at the time evolution of a single vertex, we have to minimize the interference from other vortices. So we set the block size to be much larger than the healing length $A\gg \xi$. In short time, each block can be considered as effectively isolated from others, so its evolution is almost independent. In Fig. \ref{fig:goto-static}, the left, middle and right panels show the time evolution of the density profile of the field for winding numbers $s=1,2,3$ respectively. The initial profiles (red-dashed lines) are quite different from the static solution of Eq. (\ref{Eq:vortex-profile}) (black-solid lines), but no matter what initial density profiles are chosen, either with larger densities (upper panels) or smaller densities (lower panels), the profile functions quickly approach their corresponding static limits in a time scale of about $\hbar/\mu$. The changes of densities always accompany out-going or in-coming density waves which result in a redistribution of the OAM density because the OAM is carried by particles in a rotating BEC with fixed winding numbers.

\begin{figure}[ht] 
\centering
\includegraphics[scale=0.75]{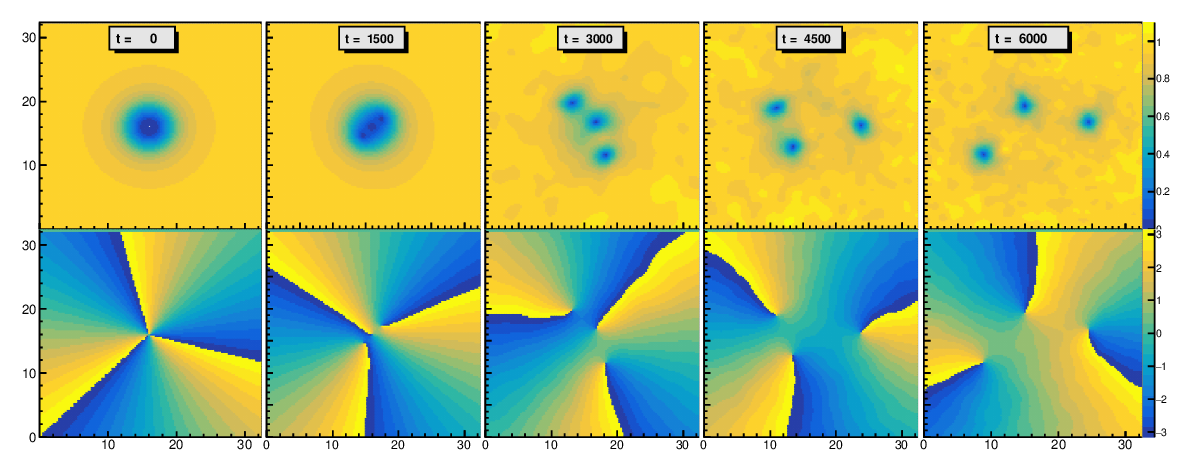}
\caption{Snapshots of the field during the decay process of an $s=3$ vortex into three primary vortices at different time in the unit of $\hbar/\mu$. Upper panels show normalized amplitudes $|\psi(\mathbf{x},t)|/\sqrt{\rho_0}$, lower panels show phases $\theta(\mathbf{x},t)$, both are functions of space coordinates in the unit of the healing length $\xi$. }
\label{fig:s3-decay-psi}
\end{figure}

\begin{figure}[ht] 
\centering
\includegraphics[scale=0.75]{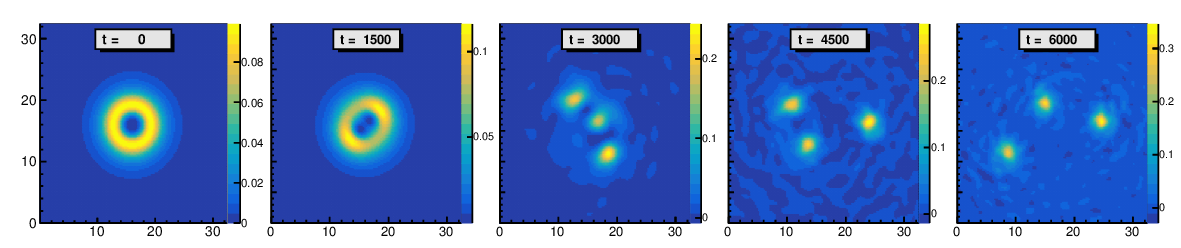}
\caption{Snapshots of the rescaled vorticity density $(\boldsymbol{\nabla} \times \boldsymbol{j})_z$ in the decay of an $s=3$ vortex into three primary vortices at different time in the unit of $\hbar/\mu$. Because the magnitudes vary significantly, the same colour indicates different values in different plots.}
\label{fig:s3-decay-vort}
\end{figure}

As shown in Eq. (\ref{vortex-energy}), the vortices with winding number $s>1$ are unstable and they will decay into primary ones with $s=1$\cite{PhysRevLett.89.190403,PhysRevA.68.023611}. The decay process takes a much longer time than the formation time $\hbar/\mu$ that we discussed in the preceding paragraph. As shown in Figs. \ref{fig:s3-decay-psi} and \ref{fig:s3-decay-vort}, the decay process of a vortex with the winding number $s=3$ starts at a time scale $\sim 10^3 \hbar/\mu $, so the life time of the metastable vortex with winding number $s>1$ is about a few thousands $\hbar/\mu $ if the vertex is put into a symmetric initial condition. The symmetric configuration at the vortex center can not resist any perturbation, which will initialize the decay of the mother vortex into $s$ primary vortices. In Fig. \ref{fig:s3-decay-psi}, we show the evolution of the field during the decay process of a $s=3$ vortex. The decay takes place when the singular point of the phase of the mother vortex splits into new singular points corresponding to a redistribution of particles' number density. As time goes on, three coupled vortices can be identified in the density profile. These vortices rotate around the center of the mother vortex in the same direction as the mother vortex. In early time, the new vortices are close to each other and prefer to be aligned in one line. Eventually, the vortices are decoupled and depart from each other with their centers forming a rotating triangle. As soon as the decoupling is completed, the density profile of each new vortex approaches the standard profile of a primary vortex. Each new vortex has deformation from the global rotation and sound-wave perturbation as the result of de-excitation. Though the interaction between the primary vortices and the sound wave is nonlinear, the primary vortices show a soliton feature in the evolution and can be considered as stable modes or new degrees of freedom.

The decay process can also be demonstrated by the kinetic energy density and vorticity. Figure \ref{fig:s3-decay-vort} shows the time evolution of the vorticity density $\mathcal{M}_{03}=(\boldsymbol{\nabla} \times \boldsymbol{j})_z$ in the decay process. $\mathcal{M}_{03}$ is re-scaled to be dimensionless so that the results can be compared directly with those in Fig. \ref{fig:stationary}. In the initial stage, the ring structure of the vortex is kept until it is perturbed and broken at $t=1500\; \hbar/\mu$. Three new vortices are formed and rotate around the center of the mother vortex as shown by the vorticity distribution. The time evolution of the kinetic energy density is similar to that of the vorticity.

\begin{figure}[ht] 
\centering
\includegraphics[scale=0.42]{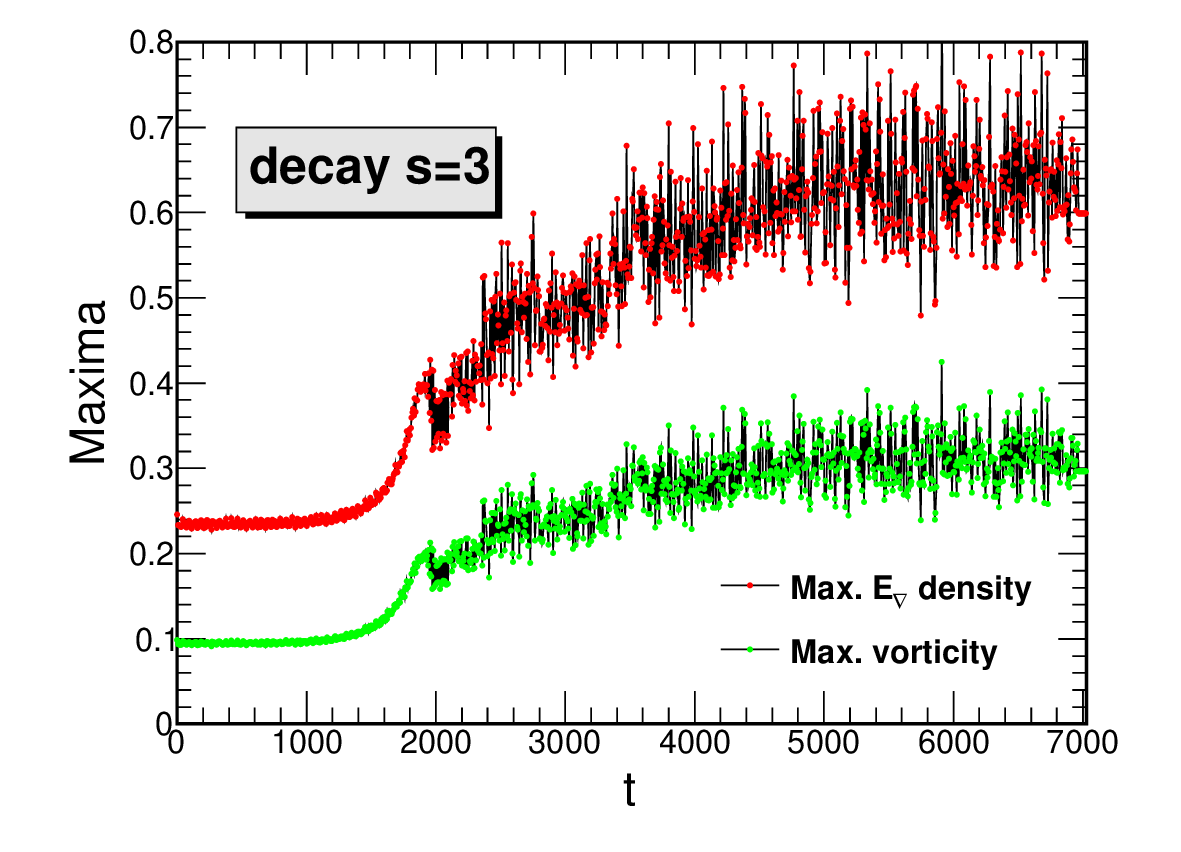}
\includegraphics[scale=0.42]{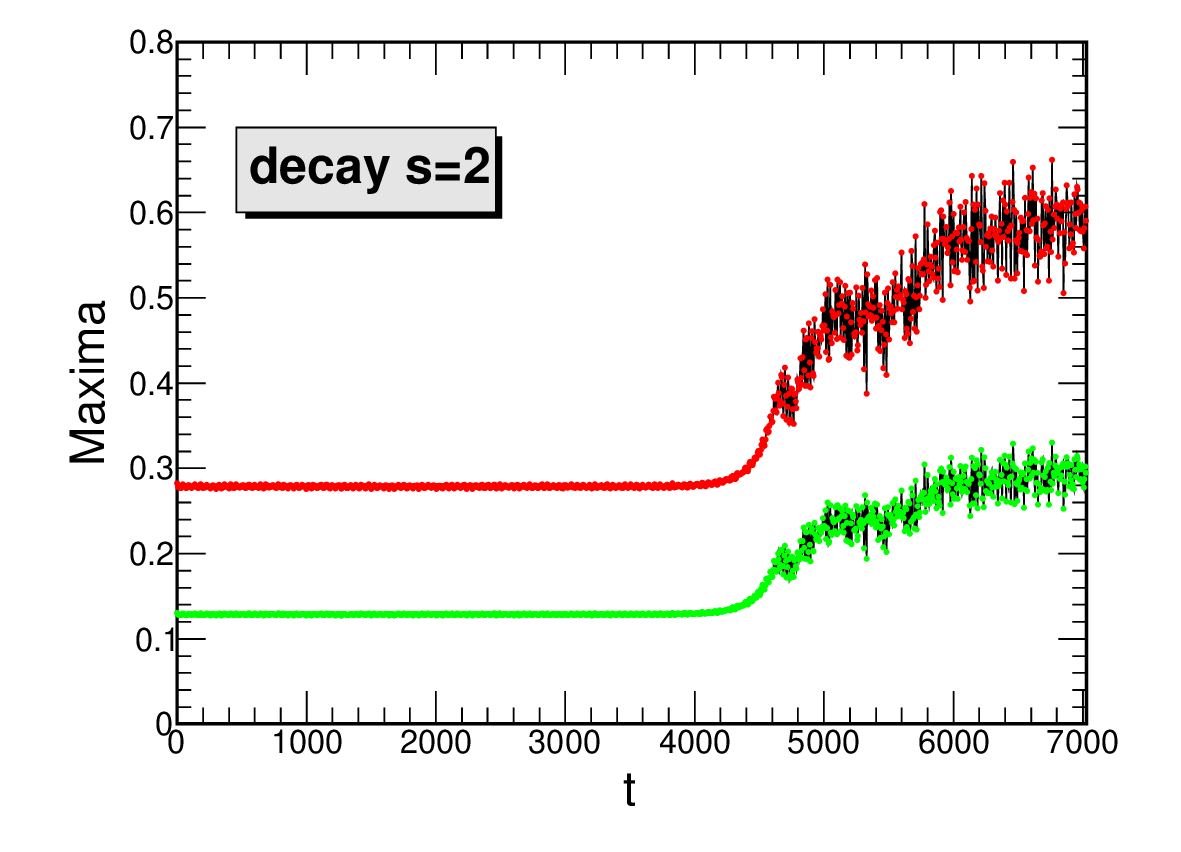}
\caption{The time evolution of the re-scaled maxima of the kinetic energy density and vorticity density during the decay. The time unit is $\hbar/\mu$. }
\label{fig:s3-decay-max}
\end{figure}

Figure \ref{fig:s3-decay-max} shows the time evolution of the rescaled maxima of the kinetic energy density and vorticity density in the decay process. One can see that the values in the early stage and later stage correspond to the mother vortex of $s=3$ and to primary vortices moving in a fluctuating environment respectively. In the early stage $t < 1000$ (time unit $\hbar/\mu $), the maxima remains stable at the values of the mother vortex. In the time interval $1000 < t < 1800$, the maxima start to increase indicating the formation of three coupled primary vortices that stay in a line and rotate around the center of the mother vortex. In the later stage $t>1800$, the fluctuations in the background become visible when the vortices start to be decoupled. The fluctuations are thought to be the results of sound waves \cite{Barenghi:2005,Suthar:2014}.

\section{Vortex production in non-central collisions}
\label{sec:collision}

\begin{figure}[ht] 
\centering
\includegraphics[scale=0.35]{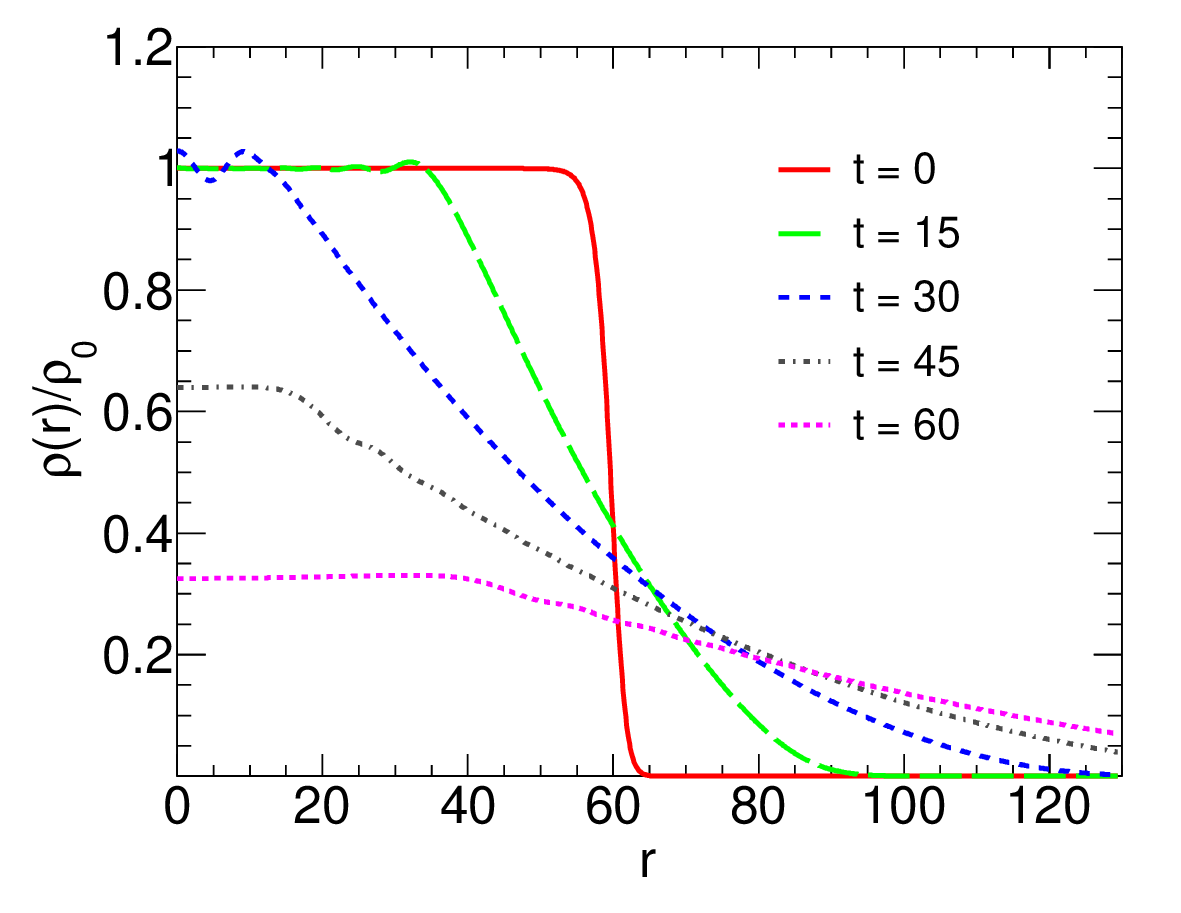}
\includegraphics[scale=0.35]{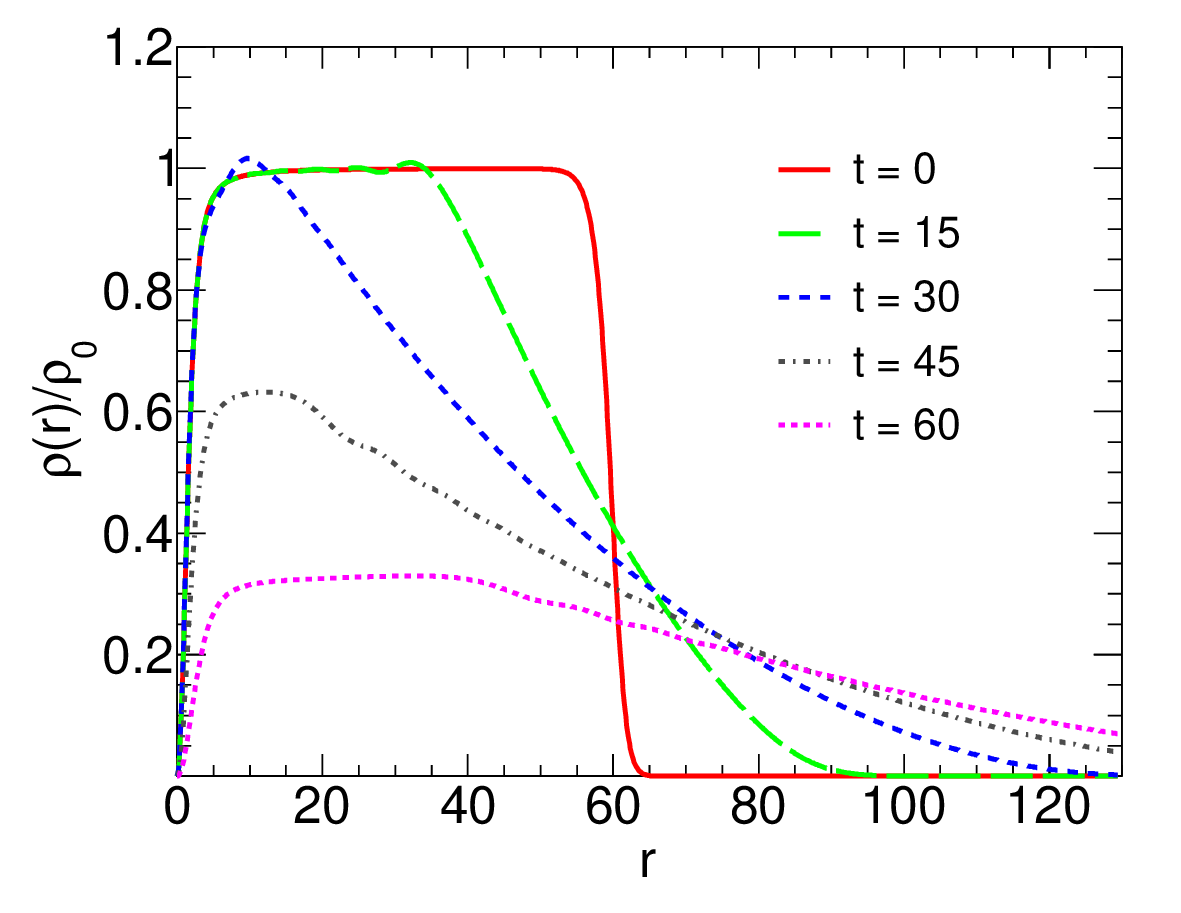}
\caption{The time evolution of the density profile $\rho(r,t)/\rho_0$ as functions of the radius in the unit of healing length $\xi$ with the time in the unit of $\hbar/\mu$. Left panel: the initial field configuration is an isolated and static disc in the Woods-Saxon distribution. Right panel: the initial field configuration is a rotating disc with a vortex at the center.}
\label{fig:free-disc}
\end{figure}

In this section, we discuss the production of vortices in non-central collisions of two BEC fields as wave packets. Experimentally, this setup can be realized by colliding two clusters of cold atoms at extremely low temperature. In our numerical simulation, we split a space area of $518\xi \times 518 \xi$ into $4096 \times 4096$ grids, each grid has a size $0.1265\xi\times 0.1265 \xi$. Similar to nuclei, we choose the initial density of the field to be in the Woods-Saxon distribution
\begin{equation}
\label{eq:Woods-Saxon}
w(r) =\frac{1}{\exp[(r-R_0)/a]+1},
\end{equation}
where $R_0$ is the radius of the disc and $a$ is the skin thickness. In our numerical calculation, we choose $R_0\approx 60\xi$ and $a\approx 1.26\xi$ to guarantee the reliability of the results. Due to the repulsive quantum pressure and self-interaction, an isolated disc is not stable on its own. This can be seen in the time evolution of the density profile in Fig.~\ref{fig:free-disc}: the field expands isotropically and the central density drops to two-thirds of its initial value at $t=45\,\hbar/\mu$.

We consider head-on collisions of two field configurations of disc shape with velocities $v_0\mathbf{e}_x$ and $-v_0\mathbf{e}_x$ at the initial time. We choose the impact parameter $(5/6)R_0$, i.e. the distance between the centers of two discs in the $y$ direction. The initial field configuration then has the form,
\begin{eqnarray}
\psi(t=0,\mathbf{r}) &=& \sqrt{\rho_0} w\left(|\mathbf{r}-\mathbf{r}_0|\right) \exp\left(\frac{i}{\sqrt{2}} \frac{v_0}{c_s} \frac{x + R_0}\xi\right) \nonumber\\
&&+  \sqrt{\rho_0} \, w \left(|\mathbf{r}+\mathbf{r}_0|\right) \exp\left(-\frac{i}{\sqrt{2}} \frac{v_0}{c_s} \frac{x - R_0}\xi\right)\,,
\end{eqnarray}
where $\mathbf{r}_0 = R_0 [ -\mathbf{e}_x + (5/12) \,\mathbf{e}_y ]$ and the factor $\sqrt{2}$ is from the definition of $c_s$. The values of $c_s$ and $\xi$ are defined at the initial density $\rho_0$.

We did the calculation with different values of the velocity $v_0$. For $v_0=0$, the two discs (field configurations) expand almost isotropically until their surfaces contact. Then interference pattern appears in the overlapping region but no vortices are generated because the system has no initial OAM. With increasing $v_0$, part of the initial OAM is deposited in an array of primordial vortices along a line separating two field configurations. The number of vortices depends on how much initial OAM is deposited into the medium. There is a vertex at the origin and other vortices locate symmetric about the origin. For small values of $v_0$, these primordial vortices expand so fast that the numerical signature of their formation is very weak. The signature becomes stronger with larger $v_0$. When $v_0$ is over some threshold, additional vortex-antivortex pairs are produced away from the array of vortices described above. In the rest of this section, we will present our results with $v_0 = 0.34\,c_s$ to demonstrate the formation of vortices followed by the production of vortex-antivortex pairs.

With our choices of $R_0$ and $v_0$, the two fields collide before significant expansion. After the collision, the system expands due to the repulsive quantum pressure and the self-interaction. There is a net radial flow of fields from the collision region to infinity, together with the produced primordial vortices. In the simulation, the periodic boundary condition is automatically guaranteed before the time when the outgoing wave front arrives at the boundary. We take a very large lattice, so that the formation of vortices and vortex-antivortex pairs in the central region is not affected by the boundary condition.

\subsection{Density distribution}

\begin{figure}[ht] 
\centering
\includegraphics[scale=0.90]{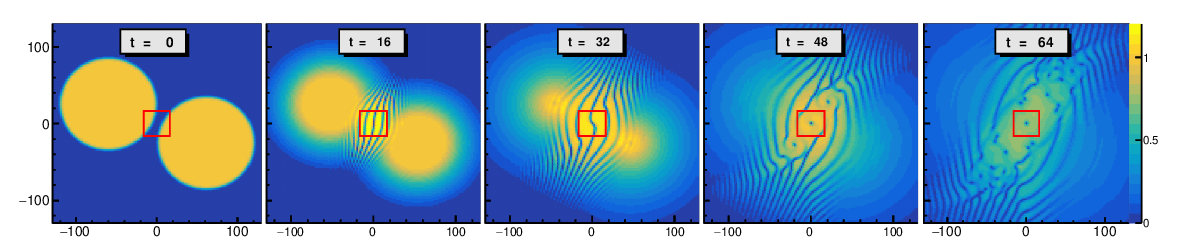}
\includegraphics[scale=0.90]{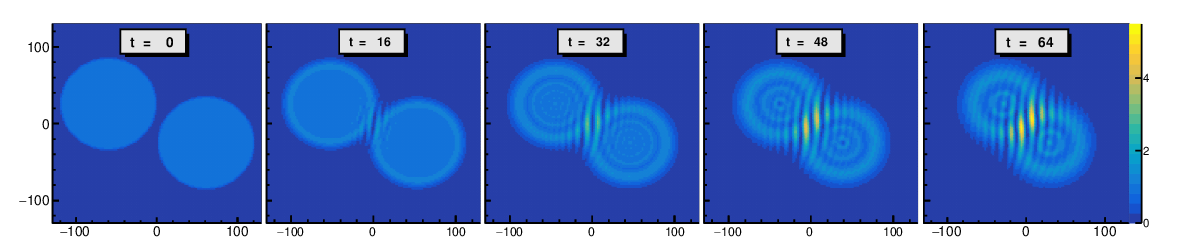}
\includegraphics[scale=0.90]{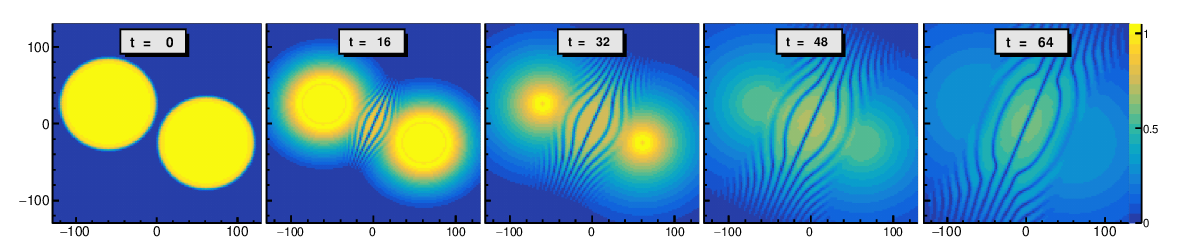}
\caption{The density distribution $\rho(x,y,t)/\rho_0$ in non-central collisions of two wave packets. The coordinates are in the unit of the healing length $\xi$ and the time is in the unit of $\hbar/\mu$. The first row: with the interaction and the initial velocity $v_0 = 0.34 c_s$. The second row: without the interaction by setting $\lambda = 0$ but with the initial velocity $v_0 = 0.34 c_s$. The third row: without initial velocity by setting $v_0 = 0$ but with the interaction.}
\label{fig:coll-I}
\end{figure}

Because of the quantum nature, two colliding wave packets interfere when they approach to each other. The interference fringes in the number density distribution are shown in Fig. \ref{fig:coll-I}. The peak density in the constructive fringes is larger than the initial density, and the density enhancement is positively correlated with the collision velocity. At the present velocity $v_0=0.34 c_s$, the enhancement factor is about $1.2$. The average density starts to decrease due to the expansion driven by the classical and quantum pressure. The purpose to choose a larger disc radius for colliding wave packets is to maintain the density in the central region for a longer time before a substantial expansion takes place.

Because of the finite OAM in the collision, the interference fringes spiral around the symmetric center, see the upper row of Fig. \ref{fig:coll-I}. Eventually, the constructive fringes merge in a larger region and tend to form a rotating area with finite size and density. The destructive fringes break near the diagonal direction into a few dots with smaller density. These dots in a rotating environment will finally become primary vortices due to self-organization. During the formation of these vortices, the density around the poles becomes symmetric, and the density profile as a function of the radius satisfies the vortex solution in a distance of a few times the healing length. To see the dynamical formation of vortices closely, we will zoom the area of the red square, where the central vortex comes into being and persists for a while before it is weakened by the final expansion.

The formation of vortices is a natural result of the velocity shear in non-central collisions. A row of vortices are generated near the diagonal direction in the overlapping region. Given the size of initial wave packets and the interaction strength, the number of vortices and the pattern in which they are aligned depend on the colliding velocity. In the current system the repulsive interaction plays the leading role in vortex formation. If we turn off the interaction by setting $\lambda=0$ in the GPE as in the second row of Fig. \ref{fig:coll-I}, the colliding wave packets just pass through each other and interfere without forming vortices in the overlapping region. If we turn off the colliding velocity of initial wave packets as in the third row of Fig. \ref{fig:coll-I}, only the expansion takes place and no vortices are formed either.

\subsection{Conservation of the winding number}

During the formation of vortices, the velocity field and the density distribution of the particle number, energy, OAM and vorticity change dramatically, but the total winding number as a measure of the topological defect is conserved. The winding number at each point is determined by the winding integral of the local velocity as,
\begin{equation}
s = \lim_{r\rightarrow 0^+}\frac{m}{2\pi  \hbar} \oint \mathbf{v} \cdot d\mathbf{r} = \lim_{r\rightarrow 0^+}\frac{1}{2\pi } \oint \boldsymbol{\nabla} \theta \cdot d\mathbf{r} =  \lim_{r\rightarrow 0^+}\frac{1}{2\pi } \int \left[\boldsymbol{\nabla} \times (\boldsymbol{\nabla} \theta)\right]\cdot d^2\mathbf{r}. \label{windnumber-1}
\end{equation}
It takes a nonzero integer when a defect is present inside the area of the integral. The sign of the integer is related to the local vorticity. The position is controlled by the surrounding environment.

\begin{figure}[ht] 
\centering
\includegraphics[scale=0.90]{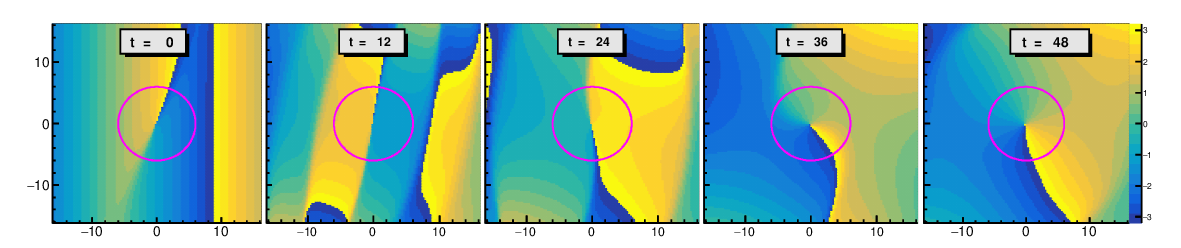}
\caption{The phase field $\theta(x,y,t)$ in red squares in Fig.\ref{fig:coll-I}. There is a defect at the center of the purple circle with the winding number $-1$ corresponding to the center vortex. }
\label{fig:coll-I-phase}
\end{figure}

As shown in purple circles in Fig. \ref{fig:coll-I-phase}, there is a defect in the symmetry center of the collision. The position of the defect is fixed and does not move with time because of the symmetric initial condition. Its winding number is always $-1$ even when the surrounding densities change dramatically. We will focus on this defect and study the dynamical formation of vortices in later subsections.

\subsection{Concentration of energy and vorticity}

\begin{figure}[ht] 
\centering
\includegraphics[scale=0.90]{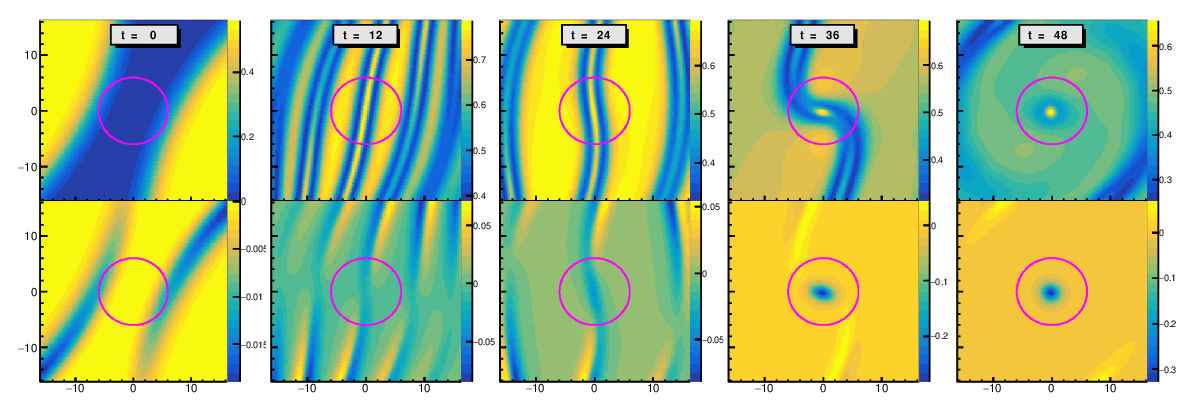}
\caption{The time evolution of the rescaled (dimensionless) energy density $\mathcal{H}/ (\lambda \rho_0^2)$ (upper panels) and the $z-$component of the rescaled (dimensionless) vorticity density $i\hbar [\boldsymbol{\nabla} \psi \times \boldsymbol{\nabla} \psi^*]_z/ ( 4m\lambda \rho_0^2/\hbar)$  (lower panels) in the red square region in Fig. \ref{fig:coll-I}. The rescale factors are chosen so that the value read from the plots can be compared directly with the results in Fig. \ref{fig:stationary}.}
\label{fig:coll-zoomin-I}
\end{figure}

During the formation of a vortex, the energy and vorticity are being concentrated towards the position of the topological defect or the vortex. Fig. \ref{fig:coll-zoomin-I} shows the time evolution of the energy density and vorticity density distributions in the central region as marked by the red square in Fig. \ref{fig:coll-I}. At $t=0$, the main contributions to the energy density are the kinetic and interaction parts, the ratio between them is a constant $v_0^2/c_s^2$. The quantum pressure involving $\mathbf{K}$ is negligible because the density changes only in the boundary region of each wave packet and it varies smoothly. Similarly the nonzero contribution of the vorticity density $\rho \mathbf{K}\times \mathbf{v}$ (we only consider its $z$ component in 2D) comes also from the boundary region. The total initial vorticity is zero, but the positive and negative contribution are separated in space according to the direction of the vector $\mathbf{K}$.

At a later time, there is a dramatic redistribution of conserved quantities, such as the particle number, energy, OAM and vorticity in the overlapping region. At $t=12$, Fig. \ref{fig:coll-zoomin-I} shows richer structures than the number density in Fig. \ref{fig:coll-I}. There is a distinct belt of the energy density in the center of the destructive fringes where the number density is small but $\theta$ and $\rho$ change fast on the belt, so that the magnitudes of $|\mathbf{v}|$ and $|\mathbf{K}|$ are large. The separation of the positive and negative vorticity is visible: the negative vorticity region is connected along the belt covering the symmetry center while the positive regions are disconnected. The difference becomes more obvious at $t=24$. As time goes on, the energy density and the negative vorticity density are being concentrated towards the symmetry center marking the formation of a vortex. At $t=48$, the formation of the vortex is almost completed with its profiles in the energy and vorticity density satisfying those of the vortex solution in Fig. \ref{fig:stationary}, while the bulk value of the number density in the vicinity is about $\rho/\rho_0\sim 1.0$.

\subsection{Evolution of average densities of particle number, OAM and vorticity}
To see how conserved quantities evolve to values of the vortex, we can look at the time evolution of the average densities of these conserved quantities. The evolution is governed by conservation laws Eqs. (\ref{N-con}, \ref{E-con}, \ref{P-con},\ref{O-con}). We will focus on the areas inside the purple circles with the radius $R=6\xi$ in Fig. \ref{fig:coll-zoomin-I} and look at the changes of conserved quantities.

\begin{figure}[ht] 
\centering
\includegraphics[scale=0.55]{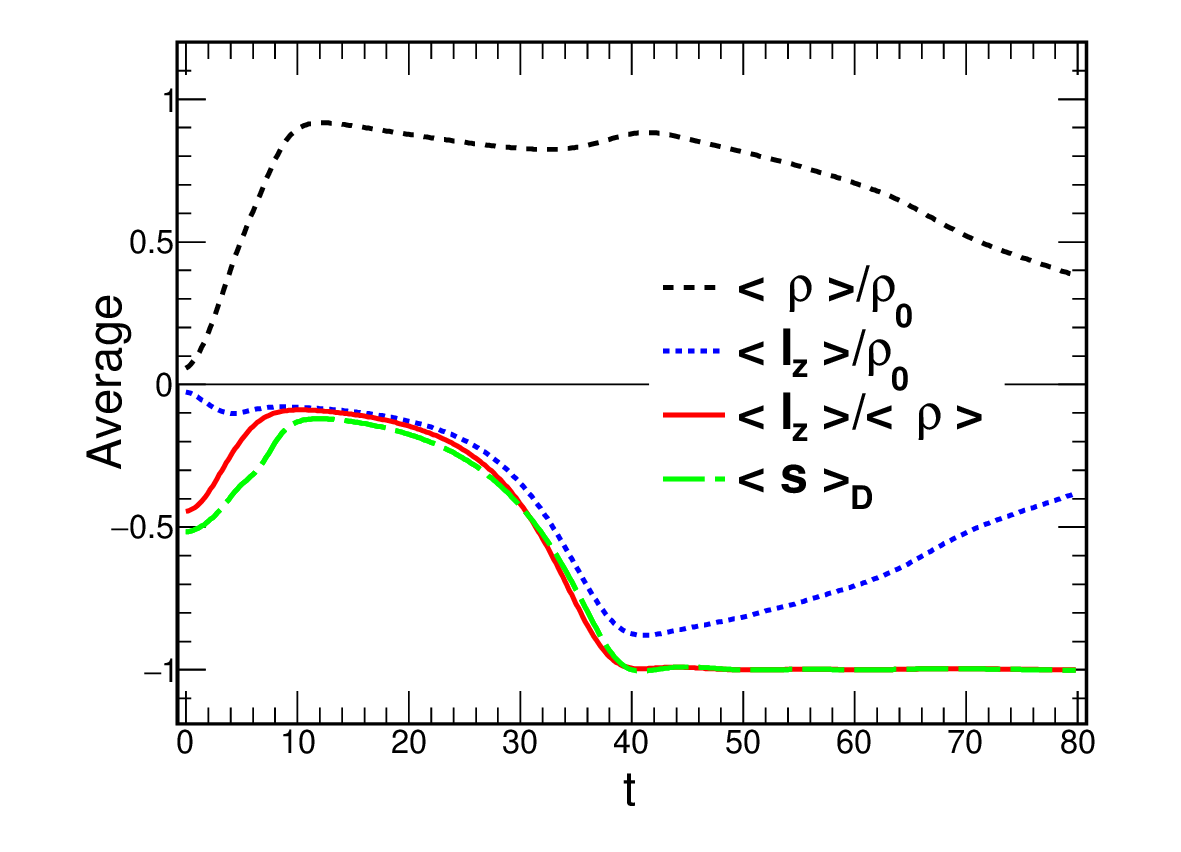}
\caption{Average densities of the particle number and OAM and the average winding number as functions of time in unit of $\hbar/\mu$.}
\label{fig:aver-OAM-I}
\end{figure}

Figure \ref{fig:aver-OAM-I} shows the average quantities as functions of time. The average number density and the $z$ component of the OAM density inside the circle are given by
\begin{eqnarray}
\langle\rho\rangle &=& \frac{1}{\pi R^2}\int d^2 \mathbf{r} \rho(\mathbf{r}) \,,\nonumber\\
\langle\boldsymbol{l}_z\rangle &=& \frac{1}{\pi R^2}\int d^2 \mathbf{r} (\mathbf{r}\times m\boldsymbol{j})_z \,.
\end{eqnarray}
The winding number weighted by the number density is defined as
\begin{equation}
\langle s \rangle _{\mathrm{D}} =  \frac{\int d\phi \rho(R,\phi) \partial \theta/\partial \phi }{\int d\phi \rho(R,\phi)} = \frac{m}{\hbar}\frac{\oint  \rho(R,\phi) \mathbf{v}(R,\phi)\cdot d\mathbf{r}  }{\int d\phi \rho(R,\phi)} = \frac{1}{2\pi \hbar} \frac{\int d^2 \mathbf{r} \left[\boldsymbol{\nabla} \times m\boldsymbol{j}(\mathbf{r})\right]_z}{\langle \rho(R) \rangle } , \label{windnumber-2}
\end{equation}
which is also the ratio of the total vorticity inside the circle to the average number density on the boundary times $2 \pi \hbar$. We see that $\langle s \rangle _{\mathrm{D}}$ matches $s$ in Eq. \eqref{windnumber-1} for a density distribution with the rotation symmetry.

We see that the number density (black-dashed line) increases, saturates, and then decreases as time goes on. In the time interval $0 < t < 10$ the number density grows linearly with time as the result of injection from a constant flow in the $x$ direction. As the density increases a pressure is formed to resist the inflow in the $x$ direction and develops an outflow in the $y$ direction. So the average number density saturates at $t=10$, then it decreases slowly in the interval $10 < t < 32$. It is interesting to notice that there is a peak around $t=40$, which comes from the breaking of the destructive interference fringe at the boundary by filling up particle density in the lower density region except the center of the vertex. After that, the average number density decreases continuously due to global expansion.

The time evolution of the average OAM inside the circle marked in Fig. \ref{fig:coll-zoomin-I} is shown by the blue-dotted line. We see that its magnitude increases in $t<5$ due to the asymmetric inflow of particles from the positive and negative $x$ direction. This part of the OAM comes from the initial geometry. During $5< t < 10$ the inflow of particles from both directions becomes almost symmetric, the average OAM is almost a constant. When the dynamical balance between the pressure and the flow is constructed, there is an exponential increase in the magnitude of the OAM during $10< t <40$. The exponential increase stops when the destructive interference fringe breaks at about $t=40$ when the magnitude reaches a maximum. At the same time, the OAM per particle $\langle\boldsymbol{l}_z\rangle/\langle\rho\rangle$ (solid-red line) saturates at the value 1 which indicates that each particle carries one unit $\hbar$ on average. After $t=40$ the OAM per particle remains a constant even though the particle number and OAM densities decrease continuously in global expansion.

The green-long-dashed line shows the average winding number in the circle as a function of time which has a similar feature to the average OAM per particle. Both indicate the collective motion inside the circle that leads to the formation of a vortex. We have tested the saturation to 1 for both the average winding number and the average OAM per particle with different radii smaller than a half distance between two topological defects. The saturation in a smaller circle is fulfilled at an earlier time, which means the formation of a vortex starts from the topological defect. As time goes on, more and more particles are involved in a vortical rotation, the effective size of the vortex increases with a rate determined by the particle number density and OAM current nearby.

\subsection{Vortex-antivortex pair production}

\begin{figure}[ht] 
\centering
\includegraphics[scale=0.76]{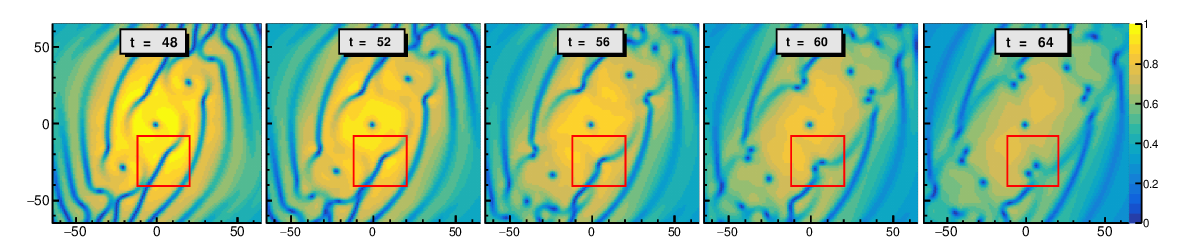}
\includegraphics[scale=0.76]{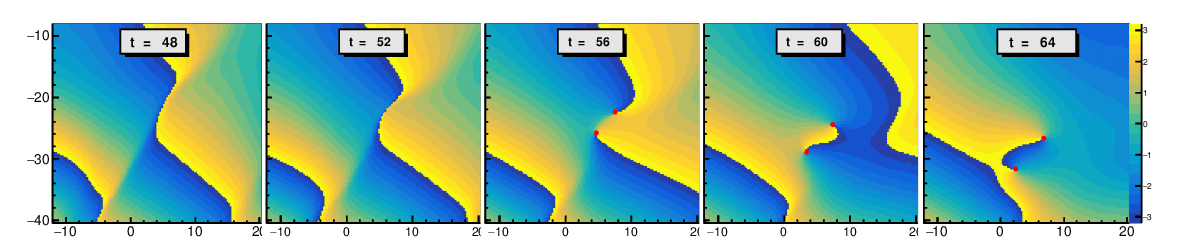}
\includegraphics[scale=0.76]{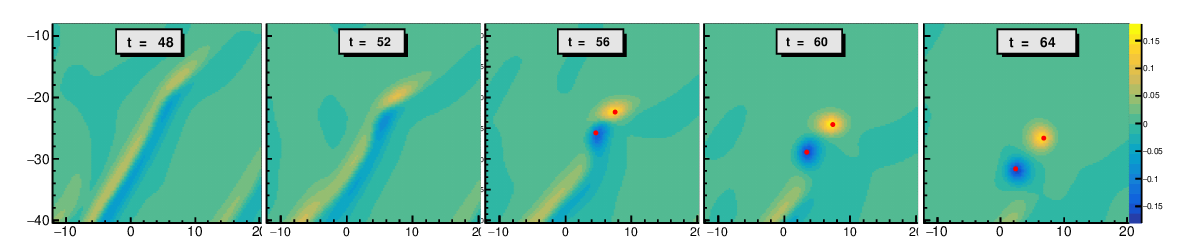}
\caption{Upper panel: the density distribution $\rho(x,y,t)/\rho_0$ as functions of coordinates in the unit of the healing length $\xi$ at different time in the unit of $\hbar/\mu$. The red squares show the production region of the vortex-antivortex pair. Middle panel: the phase function $\theta(x,y,t)$ in the red squares. Lower panel: the $z-$component of the rescaled (dimensionless) vorticity density $i\hbar [\boldsymbol{\nabla} \psi \times \boldsymbol{\nabla} \psi^*]_z/ ( 4m\lambda \rho_0^2/\hbar)$ in the red squares. In the middle and lower panel, the ends of the cut line are shown by red points, which are also the concentration positions of the vorticity density.}
\label{fig:coll-II-1}
\end{figure}

During the dynamical formation of vortices in non-central collisions, the excess energy is released by emission of sound waves. The front of an outgoing sound wave will generate a radial flow and interfere with the remnant initial flow just outside the formation region. As a consequence, the destructive fringe start to break. As shown in Fig. \ref{fig:coll-II-1}, two points with vanishing number density come into being in the broken area. At the same time, a cut line in the complex phase is generated, as shown in the middle panel. At each end of the cut line (red point in the figure), the number density is zero while the winding number has a different sign. The pair production of a positive and a negative winding number at the same time guarantees the required conservation of the winding number. Eventually, a pair of vortex-antivortex is formed around two ends at which topological defects are located.

After the formation, the vortex-antivortex pair departs in global evolution. Different from a static vortex, the vortex-antivortex pair carries a momentum which can be considered as the localized radial flow. The flow between each pair has a direction almost perpendicular to that of the flow outside.

\subsection{Remarks about simulation}

To check the precision of the numerical algorithm, for each final state of the collision a time-reversal evolution has been performed, and the initial field configuration can be reproduced correctly. We have also inverted the direction of the particle's velocity by taking the complex conjugate of the final state field, from which a proper time evolution can also reproduce the initial field configuration. This indicates that the numerical viscosity is under control and is negligible in the process of vortex production.

The production of vortices and pairs of vortices are very sensitive to the initial velocity. The larger the initial velocity is the more vortices and pairs of vortices are created in a shorter time. For smaller initial velocities, the formation of vortices depends on the competition between rotation and expansion. If the initial disc is big enough to suppress the expansion for a long time, the central vortex is more likely to be observable, because the central topological defect need more time to absorb enough OAM from the outer environment. With smaller initial velocities and a smaller disc, it is harder to form a notable vortex due to the lack of enough time and OAM.

We have tested the situation with opposite initial OAM by inverting the impact parameter of the collision. In this case, the winding numbers of single vortices (not vortex-antivortex pairs) are flipped. In central collisions with zero OAM in the initial state, there are no single vortices, but the vortex-antivortex pairs can be created if the colliding velocity is large enough.

\section{Conclusions and discussions}
\label{sec:conclusion}

We present a different approach to spin polarization from conventional
ones through quantum vortex formation in collisions of BEC. This approach
is based on the observation that the vortex is a topological excitation
in a superfluid in presence of local orbital angular momentum and
is an analogue of spin degrees of freedom. Therefore the formation
process of quantum vorticies in collisions of BEC may shed light on
the nature of the particle's spin as well as spin-orbit and spin-vorticity
couplings in the strong-interaction matter.

We have investigated the stationary property and the dynamical formation
of vortices (including vortex-antivortex pairs) by numerically solving
the GPE with a large-scale parallel algorithm on GPU to extremely
high precision. We have demonstrated that the primary vortices can
be considered as basic degrees of freedom in a system with a sizable
global OAM in the initial state. The OAM is present in a rotational
BEC or in non-central collisions of two BEC fields. The energy and
vorticity density will nucleate around topological defects, leading
to the formation of vortices in short time.

In a rotating environment, each particle carries the same local OAM, no matter how far it is from the rotation axis. An isolated vortex with a given winding number prefers to keep its own stationary properties. The density distribution of the vortex around its center responds quickly against perturbation, trying to restore its stationary profile. Even for a vortex with winding number $|s|>1$, the stationary density profile can also be restored in short time. But in long time, a $s>1$ vortex is unstable and subject to decay into primary vortices ($|s|=1$) which are the lowest energy states. The decay can be seen in the splitting of a topological defect. After that, all primary vortices are being decoupled while rotating around each other, and finally form an array of vortices. The decay precess takes much longer time than the relaxation time in the profile-restoration process.

We have observed the formation of vortices in non-central collisions of two BEC discs. A global OAM is present in the initial field configuration. If the initial global OAM is large enough, vortices are generated. The primary vortices are first generated in the central region of the collision as the result of a part of OAM is transferred to vortices. The winding number of each topological defect is conserved in the field evolution. Around a topological defect, the densities of the particle number, energy and vorticity are tuned to match the profile of the primary vortex: the particle number density reaches its local minimum while densities of the energy and vorticity (absolute value) reach their local maxima. With the emission of sound waves, an array of primary vortices are finally formed. This process can also be considered as the localization of a global rotation. If observed in a large area, vortices can be considered as fundamental objects like spin degrees of freedom. Its formation can be viewed as a consequence of the spin-orbit coupling.

In the follow-up evolution, the interaction between sound waves and interference fringes may generate vortex-antivortex pairs. The creation of vortex-antivortex pairs is not sensitive to the impact parameter, but strongly depends on the collision velocity. A vortex-antivortex pair does not carry a local OAM but carries energy and momentum in a finite region. The pair creation process in collisions is a result of redistribution of local energy and momentum densities.

The decay precess breaks the time-reversibility, i.e., when the primary vortices are decoupled, the time-reversal simulation from the final state can not recover the initial state. The reason is unclear by now, because we can not distinguish quantum decoherence from the numerical viscosity. But in short time simulation, the coherence and time-reversibility of the field are maintained to a high precision because the initial state can be well recovered by the time-reversal simulation from the final state. We also checked the influence of the numerical viscosity by inverting the direction of motion in the final state followed by a proper time evolution, and found that the initial state configuration can be recovered. This means that the numerical viscosity is under control in short time evolution.

\appendix

\section{Symplectic integration algorithm}
\label{symplectic-integrator}

In this appendix we give a brief introduction to the symplectic integration algorithm to solve the GPE (\ref{GPE}). The formal solution to the GPE can be put into the form,
\begin{equation}
\psi(t+\Delta t) = \hat{U} \psi (t)\,,
\end{equation}
where $\hat{U}$ is the evolution operator depending on the coordinate and momentum operators $\hat{x}$ and $\hat{p}$ as follows,
\bea
\hat{U} = \mathcal{T} \exp \left\{ -i \int_t^{ t+ \Delta t} d t' \left[\frac{\hat{\mathbf{p}}^2(t')}{2m} + \lambda |\psi(t',\hat{\mathbf{x}})|^2\right]\right\} \,.
\eea
The formal solution is not ready for numerical implementation because there is a time ordering operator $\mathcal{T}$ and operators are not commutable. The main difficulty is due to $e^{\hat{A}+\hat{B}} \ne e^{\hat{A}}e^{\hat{B}}$, so we cannot perform each step of the evolution independently. There is a approximation called "splitting operators" to replace the coupled operators with a bunch of separated operators, which is easy to perform and can achieve a very high precision. For example, to $\mathcal{O}(\Delta t^7)$, the evolution operator $\hat{U}$ can be rewritten as,
\begin{equation}
\hat{U} = \prod_{\alpha=1}^8 \mathrm{exp} \left( -i d_{\alpha} \Delta t\lambda |\psi(t,\hat{\mathbf{x}})|^2 \right)
\mathrm{exp} \left( -i c_{\alpha} \Delta t\frac{|\hat{\mathbf{p}}|^2}{2m} \right) + \mathcal{O}(\Delta t^7)\,,
\end{equation}
where $d_{\alpha}$ and $c_{\alpha}$ are coefficients precisely designed to achieve the desired precision. Then the evolution from $t\rightarrow t+ \Delta t$ is made by a combination of operators acting on the wave function at a given time $t$. The symplectic integrator has a symmetric form, so that the time reversal symmetry can be preserved with an extraordinary precision, $\hat{U}(t\rightarrow t+\Delta t) \hat{U}(t+ \Delta t \rightarrow t) = \mathcal{O} (\Delta t ^6) $.

Given the initial wave function in coordinate space with a periodic boundary condition, we first transform it into momentum space,
\begin{equation}
\psi_{\mathbf{p}} = \Delta \mathbf{x} ^d \sum_{\mathbf{x}} \psi(\mathbf{x}) \exp (-i \mathbf{p}\cdot \mathbf{x}) \,.
\end{equation}
So the effect of the operator $\exp (-i c_{\alpha} \Delta t|\hat{\mathbf{p}}|^2/2m)$ is a modification to the phase of the wave function in momentum space,
\begin{equation}
\psi_{\mathbf{p}}\rightarrow \psi'_{\mathbf{p}}\equiv \exp \left( -i c_{\alpha} \Delta t \frac{|\mathbf{p}|^2}{2m} \right) \psi_{\mathbf{p}}\,.
\end{equation}
Then we transform the wave function back to coordinator space,
\begin{equation}
\psi'(\mathbf{x}) = \frac{1}{V} \sum_{\mathbf{p}} \psi'_{\mathbf{p}} \exp (i \mathbf{p}\cdot \mathbf{x}) \,.
\end{equation}
The effect of the operator $\exp (-i d_{\alpha} \Delta t\lambda |\psi(\hat{\mathbf{x}})|^2)$ is to modify the phase of the wave function in coordinate space,
\begin{equation}
\psi '(\mathbf{x})\rightarrow \psi''(\mathbf{x})\equiv \exp (-i d_{\alpha} \Delta t\lambda |\psi(\mathbf{x})|^2) \psi '(\mathbf{x})\,.
\end{equation}
In this way, at each time step $\Delta t$, we have to make Fourier transformation back and forth eight times. In a successive evolution over $t$, the particle number conservation will be preserved, while the energy conservation will be slightly violated at $\mathcal{O}(\Delta t^6)$. A smaller $\Delta t$ or a higher order algorithm will help to achieve a better conservation.


\begin{acknowledgments}
We thank R. Venugopalan for the initial idea to trigger this work and for insightful discussions.  J.D. is supported by  the  Natural Science Foundation of Shandong Province under Grant No.\ ZR2020MA099. Q.W.\ is supported in part by the National Natural Science Foundation of China (NSFC) under Grants No.\ 12135011, 11890713 (a subgrant of 11890710), and by the Strategic Priority Research Program of the Chinese Academy of Sciences (CAS) under Grant No.\ XDB34030102. H.Z. is supported by the National Natural Science Foundation of China (NSFC) under Grants No.\ 12075136.
\end{acknowledgments}

\bibliography{gpe-bib-file.bib}


\end{document}